# How to minimize the environmental contamination caused by hydrocarbon releases by onshore pipelines: The key role of a three-dimensional three-phase fluid flow numerical model

Alessandra Feo[1,2,*], Emanuele Scanferla[1], Fulvio Celico[1]

1 Department of Chemistry, Life Sciences and Environmental Sustainability, University of Parma, Parco Area delle Scienze, 157/A, 43124, Parma, Italy

2 INFN Gruppo collegato di Parma, Parco area delle Scienze 7/A, 43124, Parma, Italy

*e-mail address: alessandra.feo@unipr.it

## Abstract

The contamination impact and the migration of the contaminant into the surrounding environment due to the presence of a spilled oil pipeline will cause significant damage to the natural ecosystem. For this reason, it is decisive to develop a rapid response strategy that might include accurate predictions of oil migration trajectories from numerical simulation modeling. In this paper, a three-dimensional model based on a high-resolution shock-capturing conservative method to resolve the nonlinear governing partial differential equations of the migration of a spilled light nonaqueous liquid oil contaminant in a variably saturated zone is employed to investigate the migration of the oil pipeline leakage with great accuracy. The effects on the oil type density, gasoline and diesel oil, the unsaturated zone depth, its saturation, the hydraulic gradient, and the pressure oil pipeline are investigated through the temporal evolution of the contaminant migration following the saturation profiles of the three-phase fluids flow in the variably saturated zone. The calculation results indicate that the leaking oil's pressure is the parameter that significantly affects the contaminants' arrival time to the groundwater table. Also, the water saturation of the unsaturated zone influences the arrival time as the water saturation increases for a fixed depth. The unsaturated zone depth significantly influences the contaminant migration unsaturated zone. At the same time, the oil density and the hydraulic gradient have limited effects on the contaminant migration in the variably saturated zone.

## 1. Introduction

The transfer of oil and its derivates is usually obtained using pipelines that link oil fields to refineries, or cargo oil tankers to oil storage, for example. The pipelines are considered the safest and most economical solution to transfer oil products over long distances, but problems of leakage have been frequently observed (e.g., Li et al., 2016; Xinhong et al., 2018), with a negative impact on the environment (e.g., Abbas et al., 2017; Gainer et al., 2018). In fact, construction defects, third-party damage, damages caused by landslides or earthquakes (e.g., Yan et al., 2022), as well as corrosion by the transmission fluid (e.g., Nesic, 2007; Papavinasam, 2013), can cause pipeline breakage. Based on statistical analyses, the failure probability curve for oil and gas pipelines is divided into three stages (Farshad, 2011; Liu et al., 2020); (i) initial stage of pipeline production, (ii) stable operation stage of pipelines, (iii) aging stage of pipelines. During the first stage (which usually lasts from 0.5 to 2 years), there is a high probability of breakage, mainly due to problems in pipe material, construction, etc., with about five failures per 1000 km of pipeline. The process linked to the second stage usually lasts 15 to 20 years and is mainly due to corrosion and third-party damages. In this case, the number of failures is about two times per 1000 km of pipeline. The probability of pipeline failures increases significantly during the third stage, mainly due to increased wear and corrosion.

In these scenarios, several detection techniques for pipeline leaks have been developed to reduce the loss of transported products in pipelines, even though these techniques are mainly designed for water pipelines. On the whole, the different approaches can be categorised as follows (according to Liu et al., 2020): (i) automatic, semi-automatic or manual methods; e.g., (Murvay and Silea, 2012); (ii) direct or indirect methods (e.g., Folga, 2007); (iii) optical and non-optical methods (e.g., Batzias et al., 2011); (iv) hardwater- or software-based approaches (e.g., Bie et al., 2007).

Nevertheless, despite the efforts to minimize the negative effects of pipeline failures, the protection of aquifer systems impacted by oil loss can only be guaranteed by optimising the remediation actions, in terms of effectiveness and rapidity, through the effective simulation of fluid migration in the subsurface, from the pipeline failure to the groundwater.

The contamination impact and the migration of contaminants into the surrounding environment vary on the basis of both the quantity and the quality of spilled oil products. Among the possible impacts, the one related to the infiltration of non-aqueous phase liquids lighter than water (LNAPL) is of utmost importance. LNAPLs, such as crude oil and several refined products, are hydrophobic liquid organic compounds that infiltrate vertically through the unsaturated zone (after spills from a pipeline). A fraction of the same product is trapped (discontinuous phase) within the interconnected pores (or fractures) of the aquifer medium, while another fraction (continuous phase) flows in the aquifer on the top of the groundwater table. The continuous phase migrates (or can be mobilized) according to the hydraulic gradient. Some constituents (e.g., the so-called BTEX) will dissolve in groundwater to be transported downgradient with reference to the pipeline breakage. However, LNAPLs migration is difficult to predict (and simulate) due to their features and the hydraulic heterogeneities of the aquifer media.

The determination of the spilled oil contaminant migration of the continuous phase can be described using the governing equations of immiscible phase fluid flow in a porous medium. Predicting and simulating a multiphase flow of hydrocarbon fluids in the groundwater is challenging due to the nonlinear nature and coupled properties of its governing equations. Indeed, there are still many scientific challenges to accurately describing the spilled oil subsurface multiphase flow process. Predicting the contaminated range due to spilled oil requires computationally massively numerical solutions of multiphase flow equations written in terms of soil parameters which in most cases are not exactly known due to spatial heterogeneities, parameter estimation errors, etc. (Liu et al., 2020). The nonlinearities arise since the capillary pressure and permeability of each fluid phase is a function of the saturation and, due to the gravity and pressure gradients, are responsible for creating sharp (shocks) front and rarefaction, which may introduce large errors in the numerical simulations. Many numerical models have been developed in the last decades to determine the migration of the oil saturation distribution in aquifers of various geometry and subject to different boundary water flow conditions (Hayek, 2017; Hochmuth and Sunada, 1985; Pinder and Abriola, 1986; Hossain and Corapcioglu, 1988; Høst-Madsen and Høgh Jensen, 1992; Reeves and Abriola, 1994; Lagendijk et al., 2001; Hoteit and Firoozabadi, 2008; Ataie-Ashtiani and Raeesi-Ardekani, 2010; Park et al., 2011;

Samimi and Pak, 2014; Sun et al., 2019; Sun et al., 2020; Yang et al., 2015; Xu et al., 2015; Feo and Celico, 2021, 2022). Despite the efficiency of the recent numerical models for modeling subsurface multiphase flow, analytical solutions of very simplified cases might be interesting to gain some physical insight into the flow processes. Some of them describe the nonlinear diffusion convection with a power law nonlinearity (Pistiner et al., 1989; Pistiner, 2007; 2009) where the solution is characterized by a shock wave moving front or with the inclusion of the gravity effects (Pistiner et al., 1992). Or the problem of an oil-water injection into a semi-infinite reservoir is described by a power law nonlinear convection-diffusion equation (Chen, 1988), which assumes that the position of the moving front of the water saturation profile is proportional to the square root of time.

Xu et al., 2015 present a gasoline spill modeling in a soil-aquifer system similar to the one performed in this paper. The numerical simulations for the transport and biodegradation of the gasoline spill were built up by combining the Hydrocarbon Spill Screening Model (HSSM) (Weaver et al., 1995; Charbeneau et al.,1995) for the unsaturated zone and a modified MT3DMS (Modular Three-Dimensional Multispecies Transport Model) (Zheng 2009, 2010) in the saturated one. The numerical model is divided into three steps: (1) the vertical migration of gasoline in the vadose zone, (2) the spread and dissolution processes on the groundwater table, and (3) the transport and biodegradation of dissolved benzene from gasoline in the saturated zone. The first two steps, (1) and (2), are simulated using HSSM, and the last step, (3), is simulated using a modified MODFLOW/MT3DMS. Although the actual flow in the vadose zone is three-dimensional, HSSM treats migration through the vadose zone as being one-dimensional downward.

The difference between the numerical approach (Xu et al., 2015) and this paper is that the migration of the three phases through the variably saturated zone is immiscible. The effects of volatilization/biodegradation or dissolution are not considered. We considered only immiscible phase liquids flow without dissolution, and the air density is constant, and the air density is constant (volume conservation corresponds to mass conservation). This simplification is correct as far as the temperature of the air phase is almost constant, and the pressure of the air is always of the order of the atmospheric pressure (no high-pressure air bubbles in the system). The assumptions of the computation are constant density and constant viscosity in each phase. Also, the variably saturated zone is treated and resolved as a whole, and the system is not divided into zones and treated differently (Xu et al. 2015).

The purpose of this work is to investigate the effects of an oil pipeline failure in a range of hydrogeological conditions, varying the unsaturated depth zone, the water saturation of the unsaturated zone, the oil contaminant density, two different values of the hydraulic gradient, and an oil pipeline pressure using the high-resolution shock-capturing flux (HRSC) conservative method and the CactusHydro code recently introduced in (Feo and Celico, 2021, 2022). This code allows us to accurately simulate three-phase immiscible fluid flow (in this case, gasoline and oil diesel) in a porous medium, therefore minimizing the time necessary to implement the subsequent remediation actions and optimizing the type, number, and technical features of the actions. The uniqueness of this study is to reveal the oil pipeline leakage in a variably unsaturated zone and the migration characteristics under different influencing factors, some of which have not been studied before. The results of this investigation can be used as a reference point for dealing with emergency situations involving oil leaks from pipelines.

## 2. Materials and Methods

### 2.1 Conceptual model

The numerical model developed in this paper is based on a conceptual model describing the risk of leakage of an immiscible LNAPL contaminant (gasoline or diesel oil) released to the environment from an onshore oil pipeline. The oil pipeline is situated in the unsaturated zone one-meter depth from the land surface. The possible spill would migrate toward the saturated zone under the effect of the gravity force, and it would be released for 3600 sec before being noticed and subsequently stopped by the Owner of the pipe. Due to a hydraulic gradient, the immiscible contaminant would move together with the flow while it moves downward.

A three-dimensional three-phase fluid flow numerical model is set up using CactusHydro, introduced in (Feo and Celico 2021, 2022) that has been used to predict oil spill trajectories and to precisely follow the migration of the immiscible contaminant, varying the distance of the oil pipeline with respect to the groundwater table, the saturation of the unsaturated zone, oil contaminant density, and analyzing two different values of the hydraulic gradient (0.04 and 0.004). To simplify the three-dimensional numerical simulations, we assume the variably saturated zone's grid geometry to be a parallelepiped of 160 m long with an impermeable oil pipeline along this direction situated at one-meter depth from the land surface, 80 m wide, and a variably unsaturated zone ranged between 1 and 20 meters and a saturated zone of 16 m. The leakage is represented by a cube (into the pipeline) that contains a volume of immiscible contaminant and that comes out with high pressure with respect to the atmospheric pressure of the unsaturated zone, to be around $2.0 \times 10^6 Pa$.

## 2.2 Mathematical model

This section briefly describes the numerical model introduced in (Feo and Celico, 2021, 2022) and fixes the notation. To follow the migration of a three-phase fluid flow in a porous medium composed of nonaqueous (n), water (w) and air (a), and a variably saturated zone, we impose the conservation equations for the mass and momentum for each fluid and the Darcy law for each phase,

$$\frac{\partial}{\partial t}(\phi \rho_\alpha S_\alpha) = -\frac{\partial}{\partial x^i}(\rho_\alpha u^i_\alpha) + q_\alpha, \tag{1}$$

where $\alpha = (w, n, a)$ and, Darcy's velocity for each phase is given by,

$$u^i_\alpha = -\frac{k^{ij}_\alpha}{\mu_\alpha}\left(\frac{\partial p_\alpha}{\partial x^i} + \rho_\alpha g \frac{\partial z}{\partial x^i}\right). \tag{2}$$

The governing coupled partial differential equations (PDEs) for each phase fluid are given by,

$$\frac{\partial}{\partial t}(\rho_n \phi S_n) = \frac{\partial}{\partial x^i}\left[\rho_n \frac{k_{rn}}{\mu_n} k^{ij}\left(\frac{\partial p_a}{\partial x^j} + \rho_n g \frac{\partial z}{\partial x^j}\right)\right] - \frac{\partial}{\partial x^i}\left[\rho_n \frac{k_{rn}}{\mu_n} k^{ij}\left(\frac{\partial p_{can}}{\partial x^j}\right)\right] + q_n, \tag{3}$$

$$\frac{\partial}{\partial t}(\rho_w \phi S_w) = \frac{\partial}{\partial x^i}\left[\rho_w \frac{k_{rw}}{\mu_w} k^{ij}\left(\frac{\partial p_a}{\partial x^j} + \rho_w g \frac{\partial z}{\partial x^j}\right)\right] - \frac{\partial}{\partial x^i}\left[\rho_w \frac{k_{rw}}{\mu_w} k^{ij}\left(\frac{\partial p_{caw}}{\partial x^j}\right)\right] + q_w, \tag{4}$$

$$\frac{\partial}{\partial t}(\rho_a \phi S_a) = \frac{\partial}{\partial x^i}\left[\rho_a \frac{k_{ra}}{\mu_a} k^{ij}\left(\frac{\partial p_a}{\partial x^j} + \rho_a g \frac{\partial z}{\partial x^j}\right)\right] + q_a, \tag{5}$$

where $x^i = (x, y, z)$ are the spatial cartesian coordinates, and $t$ is the time coordinate, $\rho_\alpha$ is the density $\left[\frac{M}{L^3}\right]$ of each phase, $\alpha = (w, n, a)$, $\mu_\alpha$ is the dynamics viscosity $\left[\frac{M}{LT}\right]$ of phase $\alpha$, $p_\alpha$ is the phase pressure $\left[\frac{M}{T^2L}\right]$, $k_{r\alpha}$ is the dimensionless relative permeability of phase $\alpha$, $k^{ij}$ is the absolute permeability tensor $[L^2]$, $g$ denotes the gravitational acceleration $\left[\frac{L}{T^2}\right]$, $z$ is the depth $[L]$, $S_\alpha$ is the dimensionless volumetric saturation of phase $\alpha$ that satisfy the conservation relation,

$$S_w + S_n + S_a = 1, \tag{6}$$

$q_\alpha$ is the mass source/sink $\left[\frac{M}{L}\right]$, and $\phi$ the porosity. These equations are written in terms of the variables $p_a, S_n, S_w, S_a$ and $k_{rn}, k_{rw}, k_{ra}, p_{can}, p_{caw}$.

We consider the capillary pressures for the air-water phase $p_{caw} = (p_a - p_w)$, and the capillary pressure for the air-nonaqueous phase, $p_{can} = (p_a - p_n)$, where we substituted $p_w = p_a - p_{caw}$, and $p_n = p_a - p_{can}$. They are function of $p_{caw} = p_{caw}(S_n, S_w, S_a)$, and $p_{can} = p_{can}(S_n, S_w, S_a)$. This paper uses the van Genuchten model (van Genuchten, 1980, Parker et al. 1987) for the air-water and the air-nonaqueous capillary pressure and to specify the three functions, $k_{r\alpha} = k_{r\alpha}(S_n, S_w, S_a)$:

$$k_{rw} = S_{ew}^{1/2}\left[1 - \left(1 - S_{ew}^{1/m}\right)^m\right]^2, \tag{7}$$

$$k_{ra} = (1 - S_{et})^{1/2}\left(1 - S_{et}^{1/m}\right)^{2m}, \tag{8}$$

$$k_{rn} = (S_{et} - S_{ew})^{1/2}\left[\left(1 - S_{ew}^{1/m}\right)^m - \left(1 - S_{et}^{1/m}\right)^m\right]^2, \tag{9}$$

for the relative permeabilities for three-phases, where $S_{et}$ is the total effective liquid saturation defined in terms of the irreducible wetting phase saturation $S_{wir}$. In the van Genughten model the effective saturation, $S_e$, is given by, $S_e = [1 + \alpha p_c^n]^{\left(1 - \frac{1}{n}\right)}$, and $\alpha$ and $n$ are model parameters. It can be solved for $p_c$,

$$p_c = -p_{c0}\left(1 - S_e^{1/m}\right)^{1-m}, \tag{10}$$

where $n = \frac{1}{1-m}$, and $p_{c0} = \alpha^{-1}$ is the capillary pressure at $S_e = 0$. Since $p_{caw} = p_{can} + p_{cnw}$, the capillary pressures are given by,

$$p_{can} = -p_{can0}\left(1 - S_{et}^{1/m}\right)^{1-m} \tag{11}$$

$$p_{caw} = -p_{can0}\left(1 - S_{et}^{1/m}\right)^{1-m} - p_{cnw0}\left(1 - S_{ew}^{1/m}\right)^{1-m}. \tag{12}$$

The numerical method employed in this paper has been introduced by (Feo and Celico, 2021, 2022). It is based on the high-resolution shock-capturing flux (HRSC) conservative method (Kurganov and Tadmor, 2000; Lax and Wendroof, 1960; Hou and LeFloch, 1994) to follow sharp discontinuities accurately and temporal dynamics of three-phase immiscible fluid flow in a porous medium. Several validation tests were performed to verify the accuracy of the HRSC method and the CactusHydro code. They show the absence of spurious oscillations in the solution and convergence to the "weak" solution as the grid is refined. The time evolution is performed using a forward in time explicit method rather than the most used, implicit one in which the discretization is based on a "backward-in-time" evolution. That requires the time step to be sufficiently small since the method is "conditionally stable". The implicit methods, in contrast, are "unconditionally stable" but very expensive from the computational point of view and may lead to mass balance errors.

## 2.3 Hydrogeological and hydrocarbon phase parameters

First, it investigated two types of LNAPL contaminants, gasoline and diesel oil. Table 1 shows the hydrocarbon phase properties such as density, viscosity, and parameters details used in the numerical simulations for the gasoline leak. In particular, the density of the gasoline is $750 kg/m^3$, which corresponds to an LNAPL being less dense than water.

| Parameter | Symbol | Value |
|---|---|---|
| Absolute permeability | k | $2.059 \times 10^{-11} \, m^2$ |
| Rock compressibility | $c_R$ | $4.35 \times 10^{-7} Pa^{-1}$ |

| Porosity | $\phi_0$ | 0.43 |
|---|---|---|
| Water viscosity | $\mu_w$ | $10^{-3}\ kg/(ms)$ |
| Water density | $\rho_w$ | $10^3\ kg/m^3$ |
| Oil (gasoline) viscosity | $\mu_n$ | $4.5 \times 10^{-4} kg/(ms)$ |
| Oil (gasoline) density | $\rho_n$ | $750\ kg/m^3$ |
| Air viscosity | $\mu_a$ | $1.8 \times 10^{-5} kg/(ms)$ |
| Air density | $\rho_a$ | $1.225\ kg/m^3$ |
| van Genuchten | $(n, m)$ | $\left(2.68, 1 - \frac{1}{2.68}\right)$ |
| Irreducible wetting phase saturation | $S_{wir}$ | 0.045 |
| Superficial tension air-water | $\sigma_{aw}$ | $6.5 \times 10^{-2} N/m$ |
| Superficial tension nonaqueous-water | $\sigma_{nw}$ | $2.6 \times 10^{-2} N/m$ |
| Capillary pressure air-water at zero saturation | $p_{caw0}$ | $676.55\ Pa$ |
| Capillary pressure air-nonaqueous at zero saturation | $p_{can0}$ | $405.93\ Pa$ |

Table 1. Definitions of the parameters used in the numerical simulations of a gasoline spill from an oil pipeline.

The rock compressibility $c_R$ is taken from (Freeze and Cherry, 1979), considering a porous medium being sand. The porous medium's effective porosity (at a pressure of one atmosphere) is 0.43, which corresponds to clean sandy/silty sandy (Carsel and Parrish, 1988). For simplicity, we consider a unique value of the hydraulic conductivity to be $K = 6.8 \times 10^{-5} m/s$. From here, we determined the value of the absolute permeability ($k = \frac{K\ \mu_w}{\rho_w\ g}$), see Table 1, which guarantees a contaminant's leak of around $2.78 \times 10^4 kg$ in 3600 s at a pressure of around $2.0 \times 10^6 Pa$. Indeed, the absolute permeability is $k = 2.059 \times 10^{-11} m^2$ all over the grid with $k_x = k_y = k_z$.

For the numerical simulations was implemented the van Genuchten model and an irreducible wetting phase saturation of 0.045, which is compatible with porous sand (Carsel and Parish, 1988). The van Genuchten α parameter is defined as,

$$\alpha = \left(\frac{p_c}{\rho_w g}\right)^{-1}, \qquad (13)$$

where $p_c$ is the capillary pressure head. For a clean sand/silty sand, the value is given by $\alpha = 0.145\ cm^{-1} = 14.5\ m^{-1}$. From Eq. (13) we can determine the capillary pressure at zero saturation between air and water, which gives

$$p_{caw} = \frac{\rho_w g}{\alpha} = \frac{10^3 kg/m^3\, 9.81 m/s^2}{14.5\ m^{-1}} = 676.55\ Pa$$

Using the value of the superficial tension $\sigma_{nw} = 2.6 \times 10^{-2} N/m$ and $\sigma_{aw} = 6.5 \times 10^{-2} N/m$ we have that $\beta_{nw} = \frac{\sigma_{aw}}{\sigma_{nw}} = \frac{6.5 \times 10^{-2} N/m}{2.6 \times 10^{-2} N/m} = 2.5$, and $p_{cnw}(S_w) = \frac{p_{caw}}{\beta_{nw}} = 270.62\ Pa$. Then, $p_{caw} = p_{can} + p_{cnw}$. From here we get the capillary pressure at zero saturation, $p_{can} = p_{caw} - p_{cnw} = 676.55 - 270.62 = 405.93 Pa$.

Table 2 shows, instead, the hydrocarbon phase properties such as density, viscosity, and parameters details used in the numerical simulations for the diesel oil leakage. In particular, the density of the diesel oil is $830 kg/m^3$, which also corresponds to an LNAPL being less dense than water.

| Parameter | Symbol | Value |
| --- | --- | --- |
| Absolute permeability | k | $2.059 \times 10^{-11}\ m^2$ |
| Rock compressibility | $c_R$ | $4.35 \times 10^{-7} Pa^{-1}$ |
| Porosity | $\phi_0$ | 0.43 |
| Water viscosity | $\mu_w$ | $10^{-3}\ kg/(ms)$ |
| Water density | $\rho_w$ | $10^3\ kg/m^3$ |
| Oil (diesel oil) viscosity | $\mu_n$ | $3.61 \times 10^{-3} kg/(ms)$ |
| Oil (diesel oil) density | $\rho_n$ | $830\ kg/m^3$ |
| Air viscosity | $\mu_a$ | $1.8 \times 10^{-5} kg/(ms)$ |

| Air density | $\rho_a$ | $1.225\ kg/m^3$ |
|---|---|---|
| van Genuchten | $(n, m)$ | $\left(2.68, 1 - \dfrac{1}{2.68}\right)$ |
| Irreducible wetting phase saturation | $S_{wir}$ | 0.045 |
| Superficial tension air-water | $\sigma_{aw}$ | $6.5 \times 10^{-2} N/m$ |
| Superficial tension nonaqueous-water | $\sigma_{nw}$ | $3.0 \times 10^{-2} N/m$ |
| Capillary pressure air-water at zero saturation | $p_{caw0}$ | $676.55\ Pa$ |
| Capillary pressure air-nonaqueous at zero saturation | $p_{can0}$ | $374.68\ Pa$ |

Table 2. Definition of the parameters used in the numerical simulations of a diesel oil spill from an oil pipeline.

Table 1 and Table 2 differ in the value of the immiscible contaminant released. As for the capillary pressure, following the same procedure used for the gasoline case and using the value of the superficial tension of the diesel oil, $\sigma_{nw} = 3.0 \times 10^{-2} N/m$ and $\sigma_{aw} = 6.5 \times 10^{-2} N/m$. Indeed, we have that $\beta_{nw} = \dfrac{\sigma_{aw}}{\sigma_{nw}} = \dfrac{6.5 \times 10^{-2} N/m}{3.0 \times 10^{-2} N/m} = 2.17$, and $p_{cnw}(S_w) = \dfrac{p_{caw}}{\beta_{nw}} = 311.77\ Pa$. Since $p_{caw} = p_{can} + p_{cnw}$. From here we get the value of the capillary pressure at zero saturation, $p_{can} = p_{caw} - p_{cnw} = 676.55 - 311.77 = 374.68\ Pa$.

## 3. Three dimensional numerical simulations results and discussions

This section presents a selection of twelve results out of twenty-eight (see Supplementary Material) different three-dimensional numerical simulations of immiscible leaks of LNAPL (gasoline or diesel oil) that migrate in a variably saturated zone from an oil pipeline. The migration is investigated by varying: the unsaturated depth zone, the hydraulic gradient, and the water saturation of the unsaturated zone starting from dry soil porous medium. It is considered a three-phase fluid model and follows the temporal evolution of the migration of the contaminant following the saturation profiles of each phase along the variably saturated zone. We also determine the time for the LNAPL contaminant to arrive at the groundwater table from the oil pipeline and the leakage location. See Table 3 for details. Different situations are analyzed in which the distance from the contaminant and the groundwater table varies from 1 to 2, 5, 10, and 20 meters, two different values of the hydraulic gradient (0.04 and 0.004) and different levels of water saturation in the unsaturated zone (0, 0.20 and 0.50).

| $S_w$ (unsaturated zone) | 0 | 0 | 0 | 0 | 0 | 0 | 0 | 0 | 0 | 0 | 0.2 | 0.5 | 0.2 | 0.5 |
|---|---|---|---|---|---|---|---|---|---|---|---|---|---|---|
| Unsaturated zone depth | 1m | 1m | 2m | 2m | 5m | 5m | 10m | 10m | 20m | 20m | 1m | 1m | 10m | 10m |
| Hydraulic gradient | 0.04 | 0.004 | 0.04 | 0.004 | 0.04 | 0.004 | 0.04 | 0.004 | 0.04 | 0.004 | 0.04 | 0.04 | 0.04 | 0.04 |
| Gasoline | M | S | S | S | S | S | M | S | S | S | M | M | M | M |
| Diesel oil | M | S | S | S | S | S | M | S | S | S | M | M | M | M |

Table 3. Numerical simulation cases and where to find them. M stands for this paper, S stands for Supplementary material.

## 3.1 Numerical simulations of a gasoline leak from an oil pipeline

First, it is considered a leak of gasoline released into the environment from an oil pipeline situated in the unsaturated zone at a one-meter depth from the land surface for a fixed time of 3600 s. The immiscible contaminant moves inside the oil pipeline with a pressure that may vary between $(2.0 - 2.4) \times 10^6 Pa$, while the rest of the unsaturated zone is at the atmospheric pressure. After the contaminant leakage, it migrates from the unsaturated zone toward the saturated one under the effect of the gravity force. For the transient numerical simulations, we consider a three-phase fluids model composed of water, air, and a fixed quantity of gasoline. We follow the contaminant released through the variably saturated zone.

Fig 1 shows one of the grid geometries used in this paper. We assume the variably saturated zone grid geometry to be a parallelepiped of $160\ m$ long from $x = [-100, +60]m$ (left-hand side), $80\ m$ wide from $y = [-40, +40]\ m$ (right-hand side), and $18\ m$ depth from $z = [+2, -16]m$. We consider (otherwise mentioned) a spatial resolution of $dx = dy = dz = 0.50\ m$, and a time step resolution of $dt = 0.025\ s$. The leakage of contaminant is situated at $(x, y, z) = (0,0,1)\ m$. The porous medium is composed of an unsaturated dry zone (air-NAPL) and a saturated one (where initially there is only water) separated by the groundwater table. The groundwater table cross the point $(x, y, z) = (0,0,0)\ m$. All boundary conditions are no-flow, except in the infiltration zone on top of the parallelepiped. The bottom part of the parallelepiped is impermeable, with an absolute permeability of $2.059 \times 10^{-15} m^2$ (ten thousand smaller than the value of the rest of the domain, see Table 1 for details). The dimension of the grid has been chosen to cover the minimum possible part of the aquifer. However, it is big enough to avoid the boundary effect (finite grid size) on the dynamics of the contaminant.

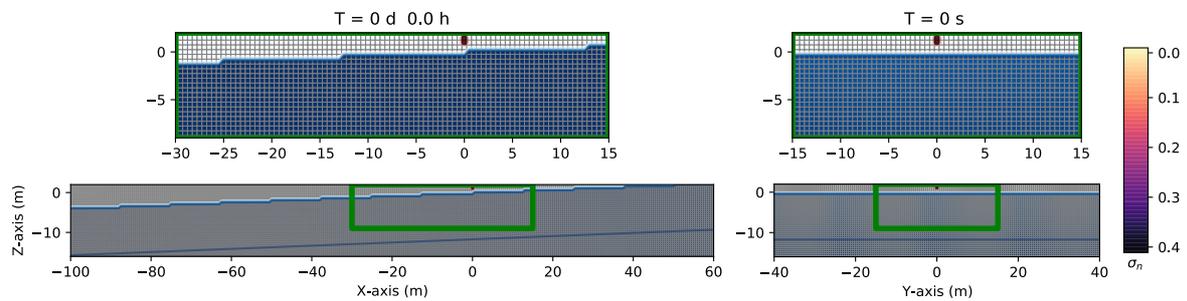

Fig 1. Example of the three-dimensional grid geometry used in the numerical simulation of a three-phase fluid flow (water + LNAPL + air) with a spatial grid resolution of $0.50\ m$ and a grid dimension of $160\ m \times 80 \times 18\ m$, at the initial time $t = 0\ s$. The immiscible contaminant is situated at the top of the parallelepiped in the $z - x$ plane (left-hand side) and the $z - y$ plane (right-hand side), respectively.

After being released, from the oil pipeline, into the environment, the LNAPL migrates downward into the unsaturated zone under the influence of gravity. The contaminant is released in all directions in the cartesian coordinates system, assuming no preferential direction, with a pressure equal to $2.0 \times 10^6 Pa$. Fig 2 shows the three-dimensional numerical simulation results of the saturation contours, $(\sigma_n = S_n \phi)$, at different times, for the three-phase immiscible fluid flow (water + gasoline + air) in a dry unsaturated zone, using a spatial grid resolution of $0.50\ m$ and a grid dimension of $160\ m\ \times\ 80\ m\ \times\ 18\ m$. The saturation contours are viewed in the $z - x$ plane on the left-hand side. On the right-hand side, instead, in the $z - y$ plane. The green rectangle zone in each panel is amplified in the upper area at each time indicated.

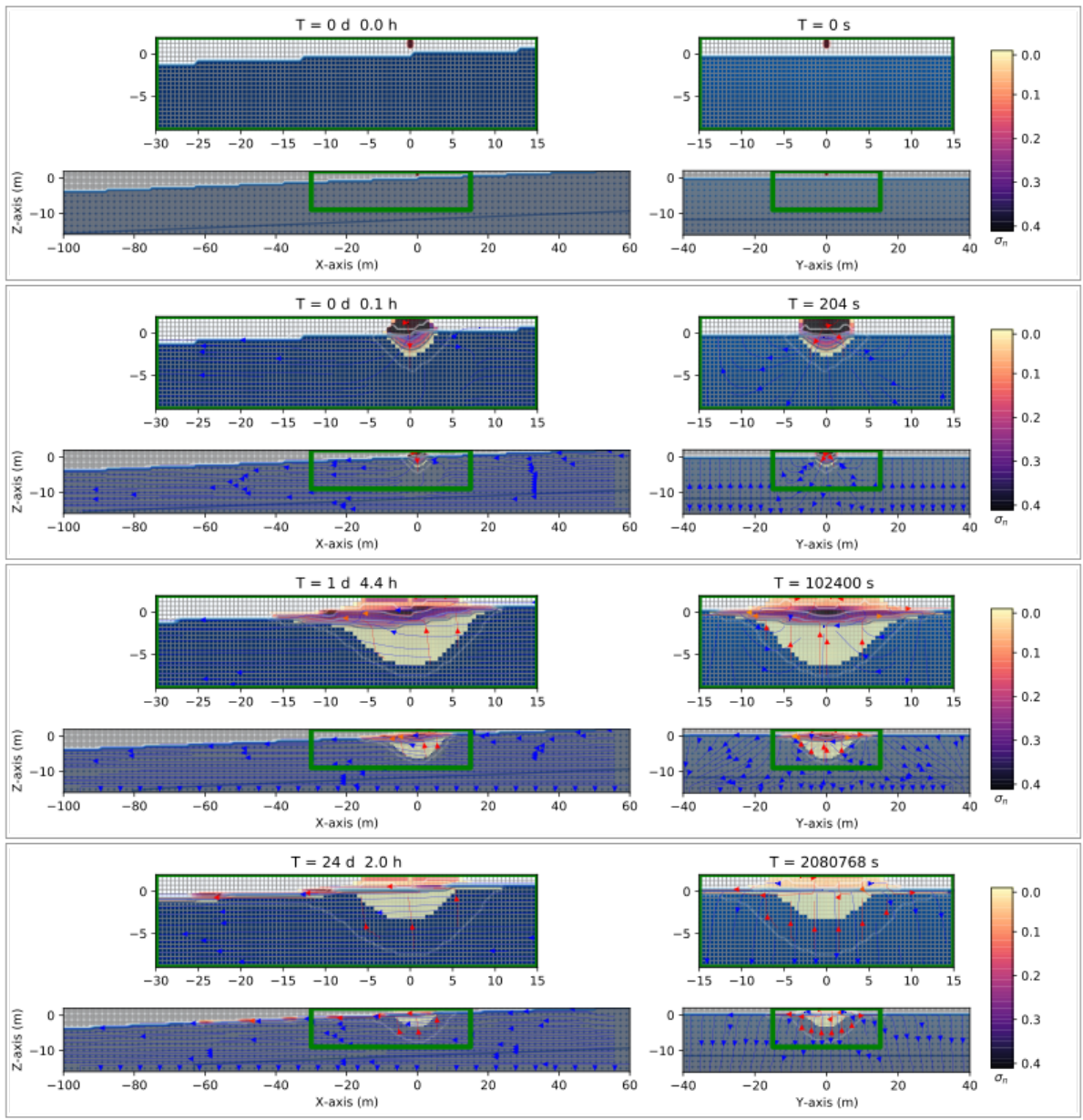

Fig 2. Three-dimensional numerical results on the saturation contours ($\sigma_n = S_n \phi$) of a three-phase immiscible fluid flow (water + gasoline + air) in a dry unsaturated zone, using a spatial grid resolution of 0.50 $m$ and a grid dimension of 160 $m$ × 80 $m$ × 18 $m$, at different times. A hydraulic gradient of 0.04. Left-hand side shows the saturation contours in the $(z - x)$ plane. Right-hand side shows the saturation contours in the $(y - x)$ one. The spill is released from an oil pipeline at $(x, y, z) = (0,0,1)\ m$.

The contaminant generally moves predominantly downward in the unsaturated zone due to the gravity force. Lateral spreading may also appear due to the presence of capillary pressures and the high pressure of the contaminant itself. Fig 2 shows the initial time, $t = 0\ s$ (the first row), and the position of the contaminant. At 204 s (the second row), the contaminant has arrived at the groundwater table. Although a consistent quantity remains in the capillary fringe, some material enters the saturated zone due to the elevated contaminant pressure. The third row shows the contaminant at one day and 4.4 hours. Notice that the leak spill from the pipeline is already stopped. The fourth row shows the fate of the immiscible contaminant, which has already arrived, and remained, in the capillary fringe being an LNAPL, moving into the left (negative) direction of the x-axis together with the water (see the left-hand side) due to a hydraulic gradient of 0.04. (See the blue arrows directed to the left-hand side). In contrast, the right-hand side remains symmetric around the plane $z - y$. After 24 days and 2.0 hours, the contaminant reaches position $x = -62\ m$.

Fig 3 shows three-dimensional numerical simulation results of the depth as a function of the water saturation $S_w$ (blue points), gasoline saturation $S_n$ (red points), and air saturation $S_a$ (green points) at various times for a gasoline leak shown in Fig 2. Initially, at $t = 0\ s$ (left-hand side), there is a sharp front of immiscible contaminant located at $z = 1.0\ m$ (red points) with a saturation $S_n = 1$. Below $z = 1.0\ m$, this saturation goes abruptly to zero since initially the contaminant is situated only on top of the parallelepiped. See also Fig 2, the first row. Instead, the water saturation (blue points) is zero in the unsaturated zone (being completely dry). It is equal to one in the saturated zone, which initially is filled up with water. The air saturation (green points) equals one in the unsaturated region with no contaminant. Notice that the sum of the three saturations is always one at any depth value, as stated in Eq. (6). After 5.7 hours, the immiscible gasoline has already reached the saturated zone. Being an LNAPL remains around the capillary fringe, but because it reaches the aquifer with a high pressure, it enters a bit into the saturated zone (center, see the red point that reaches almost a depth of 5 meters. Also compared with Fig 2, the third row). After two days and 8.9 hours, the contaminant tends to move to the left direction due to a hydraulic gradient. Also, the contaminant that initially enters the saturated zone comes back to the capillary fringe and stays in this zone (right-hand side).

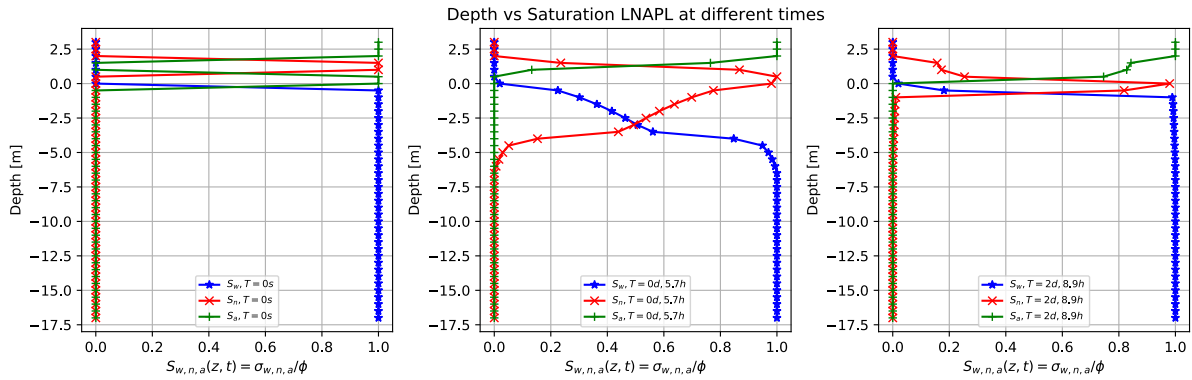

Fig 3. Three-dimensional numerical simulation results of the depth as a function of the water saturation $S_w$ (blue points), gasoline saturation $S_n$ (red points), and air saturation $S_a$ (green points) at various times for a gasoline leak of Fig. 2. Initially, at t = 0 s, there is a sharp front of contaminant saturation situated on top of the grid which rapidly goes to zero when height decreases. At the same time, it is filled by the air saturation (green point) in the unsaturated zone and the water saturation in the saturated zone. At later times the contaminant has already reached the aquifer zone, although a small part of it enters the saturated zone (center) and remains floating while moving with the direction of the groundwater flow.

Fig 4 shows the three-dimensional numerical simulation results of the saturation contours, ($\sigma_n = S_n \phi$), at different times, for the three-phase immiscible fluid flow (water + gasoline + air) using a grid resolution of $0.50\ m$ and a grid dimension of $160\ m$ long from $x = [-100, +60]m$ (left-hand side), $80\ m$ wide from $y = [-40, +40]\ m$ (right-hand side), and $27\ m$ depth from $z = [+11, -16]m$. The leakage contaminant is initially situated at $(x, y, z) = (0,0,10)\ m$ in the oil pipeline. After one hour (at 3686 s), the immiscible contaminant still has not reached the groundwater table (the first row). At 6553 s, the contaminant arrives at the groundwater table and enters the saturated zone (see the second row at 7168 s) but remains in the capillary fringe zone, being an LNAPL. Then it moves in the left direction due to a hydraulic gradient equal to 0.04 (the third row). After 12 days and 1.2 hours (the last row), the contaminant reaches the position $x = -41.0\ m$ (the fourth row). Notice that the contaminant saturation on top of the grid is decreasing since the leakage in the pipeline is closed after one hour. The number of cells used for this numerical simulation is $160\ \times 80\ \times 27\ = 345600$.

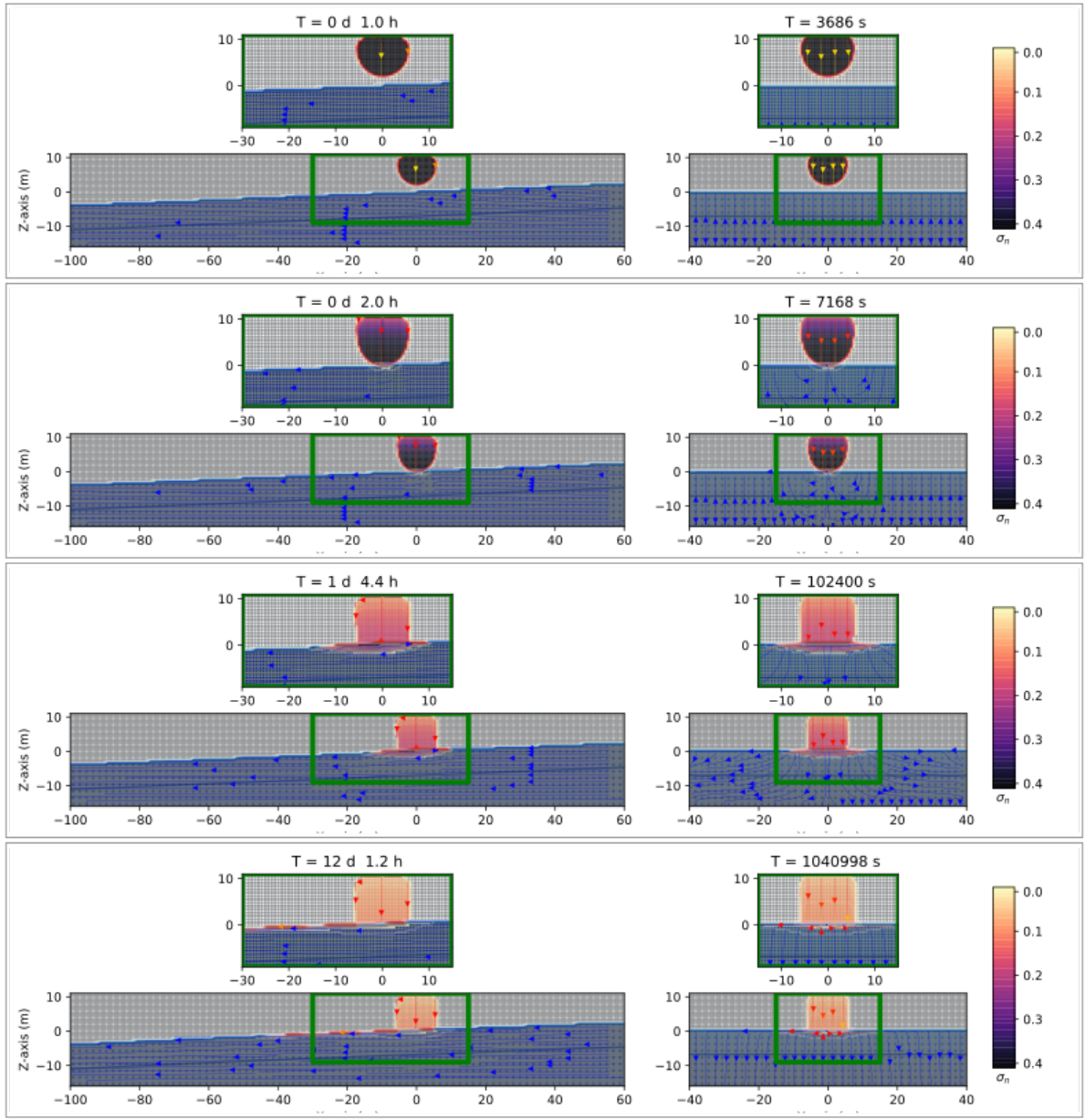

Fig 4. Three-dimensional numerical results on the saturation contours ($\sigma_n = S_n \phi$) of a three-phase immiscible fluid flow (water + gasoline + air) using a spatial grid resolution of 0.50 $m$ and a grid dimension of 160 $m$ × 80 $m$ × 27 $m$, at different times. A hydraulic gradient of 0.04. The spill is released from an oil pipeline at $(x, y, z) = (0,0,10)$ $m$. Notice that the first panel corresponds to a time equal to 3686 s rather than the initial time equal to zero.

Fig 5 shows similar three-dimensional numerical simulation results to the one in Fig 2, but the hydraulic gradient is 0.004 instead of 0.04, (see Fig 1S in the Supplementary material for details). Fig 5 shows this migration in the $y - x$ plane. It can be noticed that, after 21 d and 10.3 h, the contaminant has reached position $x = -20\ m$ (on the left-hand side of the panel, the right-hand side is its amplification). The hydraulic gradient has a key role in the movement of the contaminant in the direction of the groundwater flow. And has the consequence that the migration in the flow direction is lower when the hydraulic gradient is smaller.

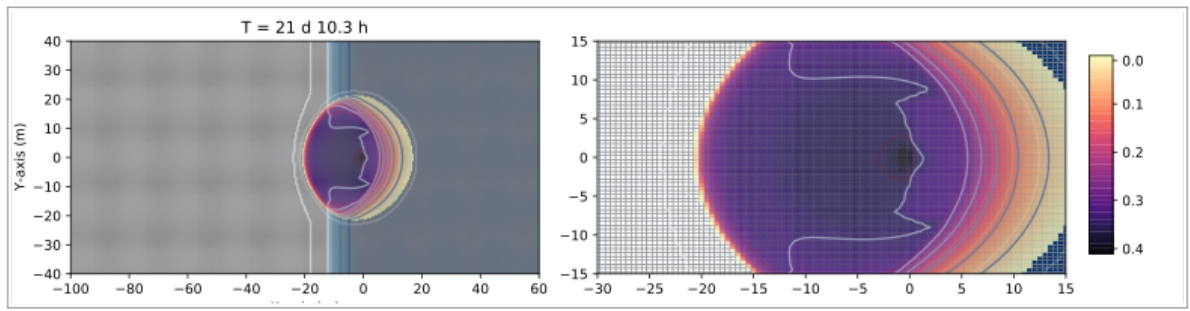

Fig 5. Three-dimensional numerical results on the saturation contours ($\sigma_n = S_n \phi$) of a three-phase immiscible fluid flow (water + gasoline + air) using a spatial grid resolution of 0.50 $m$ and a grid dimension of 160 $m$ × 80 $m$ × 18 $m$, at different times in the $y - x$ plane. A hydraulic gradient of 0.004. The spill is released from an oil pipeline at $(x, y, z) = (0,0,1)\ m$.

### 3.1.1 Numerical simulations of a gasoline leak from an oil pipeline in a partially unsaturated zone

The effects of the water saturation in the unsaturated zone are investigated using the same situation as before, but with a porous soil unsaturated zone, which is not completely dry but with nonzero water saturation. Consider two different cases, $S_w = 0.20$ and $S_w = 0.50$. This situation can be imagined as continuous rain, partially filling the unsaturated zone with water. The grid used here is like the one shown in Fig 1 (for one meter). The boundary conditions are similar to the previous cases. Except that, at each point of the grid in the unsaturated zone, $S_w$ is different from zero, and at later times on top of the grid, this value is kept constant. In this way the water saturation of the unsaturated zone remains constant.

Fig 6 shows the numerical simulation results similar to Fig 2, except that the unsaturated zone now has $S_w = 0.20$ instead of $S_w = 0.0$ (dry soil). The first difference with respect to Fig 2 is that there is a vertical water flow in the unsaturated zone (see the blue lines) due to the gravity force (beside the one going in the left direction in the saturated zone). Initially, the contaminant is released on top of the grid, and at 204 s already arrived at the groundwater table (the second row) in Fig 6. In this case, the presence of the water in the unsaturated zone does not substantially change the situation with respect to the dry case (see the third row at one day and 4.4 hours, in both figures).

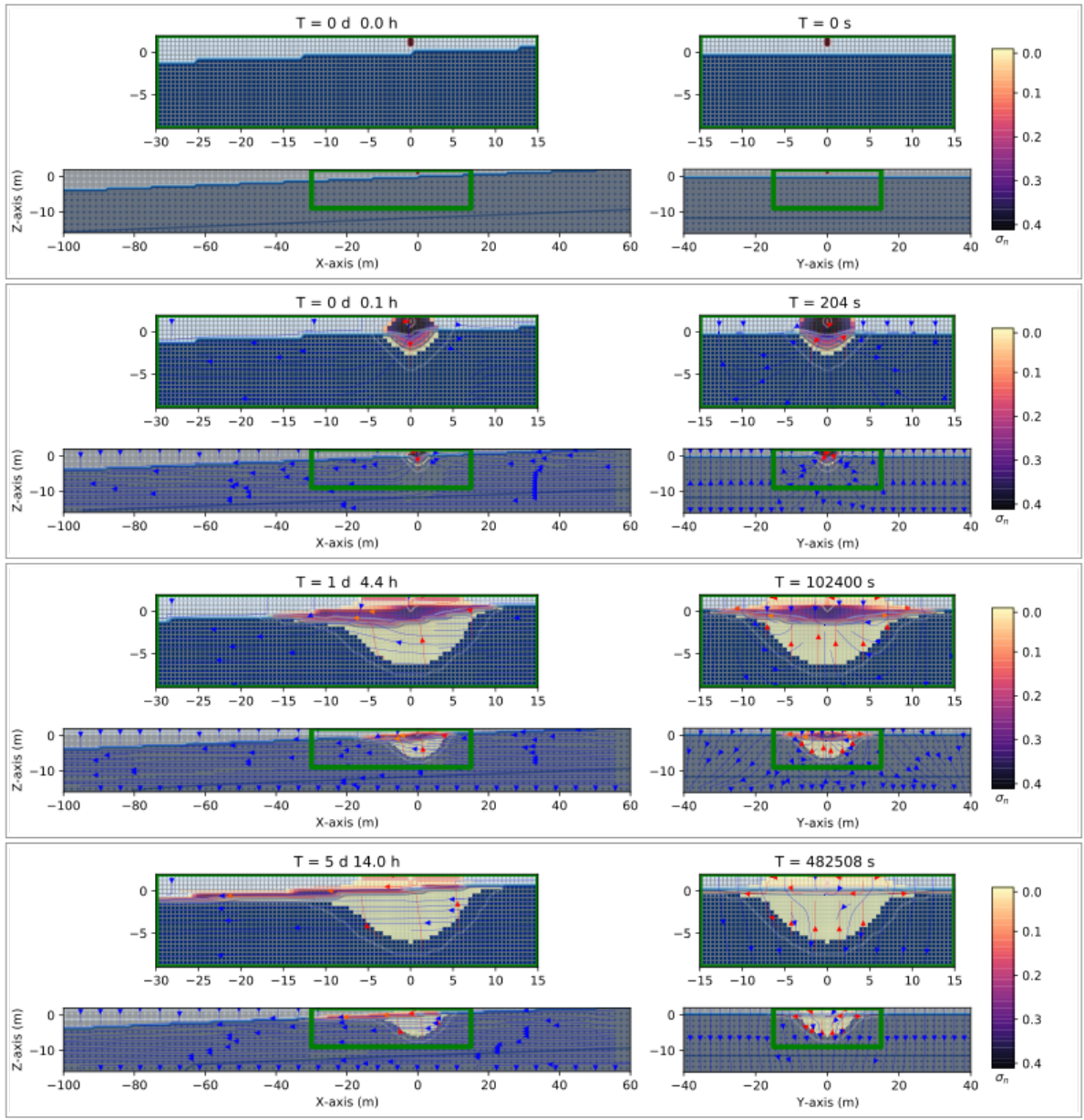

Fig 6. Three-dimensional numerical results on the saturation contours ($\sigma_n = S_n \phi$) of a three-phase immiscible fluid flow (water + gasoline + air) using a spatial grid resolution of 0.50 $m$ and a grid dimension of 160 $m$ × 80 $m$ × 18 $m$, at different times. A hydraulic gradient of 0.04. The spill is released from an oil pipeline at $(x, y, z) = (0,0,1)\ m$. The unsaturated zone has a $S_w = 0.20$.

Fig 7 shows a similar case as in Fig 6, but the water saturation in the saturated zone is increased to $S_w = 0.50$. In this case, the contaminant reaches the groundwater table before 204 s (the second row), which means there is no change with respect to the previous case due to the high pressure of the contaminant leakage. After 1 d and 4.4 hours, the level of the groundwater table increases due to a filling of the voids of the porous medium. See the third and fourth rows, respectively.

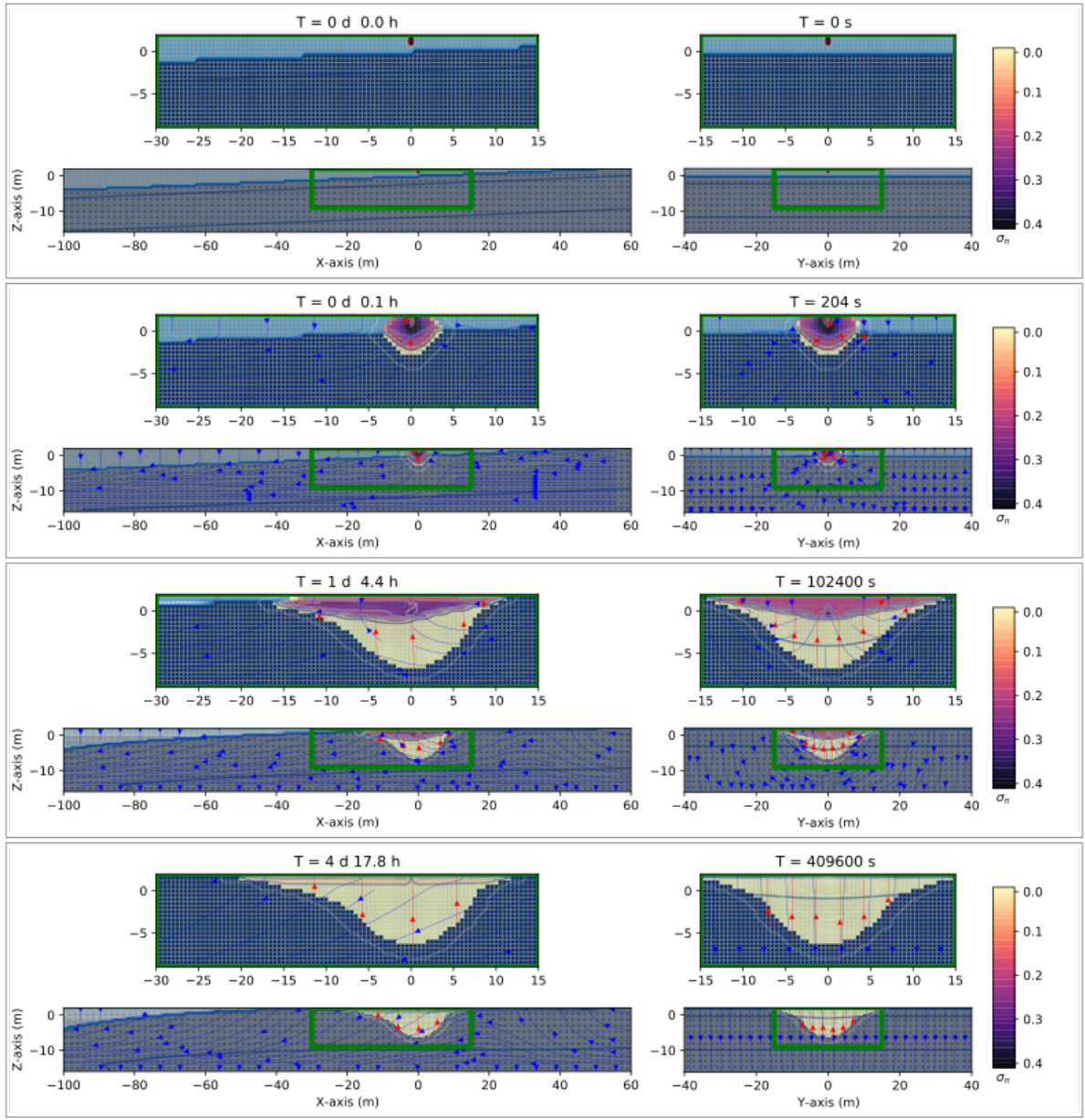

Fig 7. Three-dimensional numerical results on the saturation contours ($\sigma_n = S_n \phi$) of a three-phase immiscible fluid flow (water + gasoline + air) using a spatial grid resolution of 0.50 $m$ and a grid dimension of 160 $m$ × 80 $m$ × 18 $m$, at different times. A hydraulic gradient of 0.04. The spill is released from an oil pipeline at $(x, y, z) = (0,0,1)\ m$. The unsaturated zone has a $S_w = 0.50$.

Similar numerical simulations were performed, varying the depth of the unsaturated zone. Fig 8 shows the saturation contours ($\sigma_n = S_n \phi$), at different times. The source of contaminant is initially situated at $(x, y, z) = (0,0,10)\ m$ in the oil pipeline, and the saturation of the water in the unsaturated zone is $S_w = 0.20$. Fig 9, instead, shows a similar scenario to Fig 8, but the saturation of the water in the unsaturated zone is increased to $S_w = 0.50$. Notice in both figures the presence of vertical water flow (see the blue lines) in comparison with Fig 4, which corresponds to a dry unsaturated zone case.

The presence of water saturation in the unsaturated zone accelerates the arrival of the contaminant into the aquifer. In the case of Fig 8, the contaminant arrives at the groundwater table at 4915.0 s (see the second row), while in Fig 9 it arrives at 4710.4 s (see the second row). The more water content in the unsaturated zone, the faster the contaminant reaches the saturated zone. Also, the presence of water in the unsaturated zone inhibits the spill of contaminant from the pipeline in comparison to the dry soil case (Fig 4). Indeed, the quantity of contaminant spilled after one hour is 213904 kg in Fig 4, 197943 kg in Fig 8, and 158647 kg in Fig 9. The saturation contours values of $\sigma_n$ are lower in Fig 9 than in the case of Fig 8.

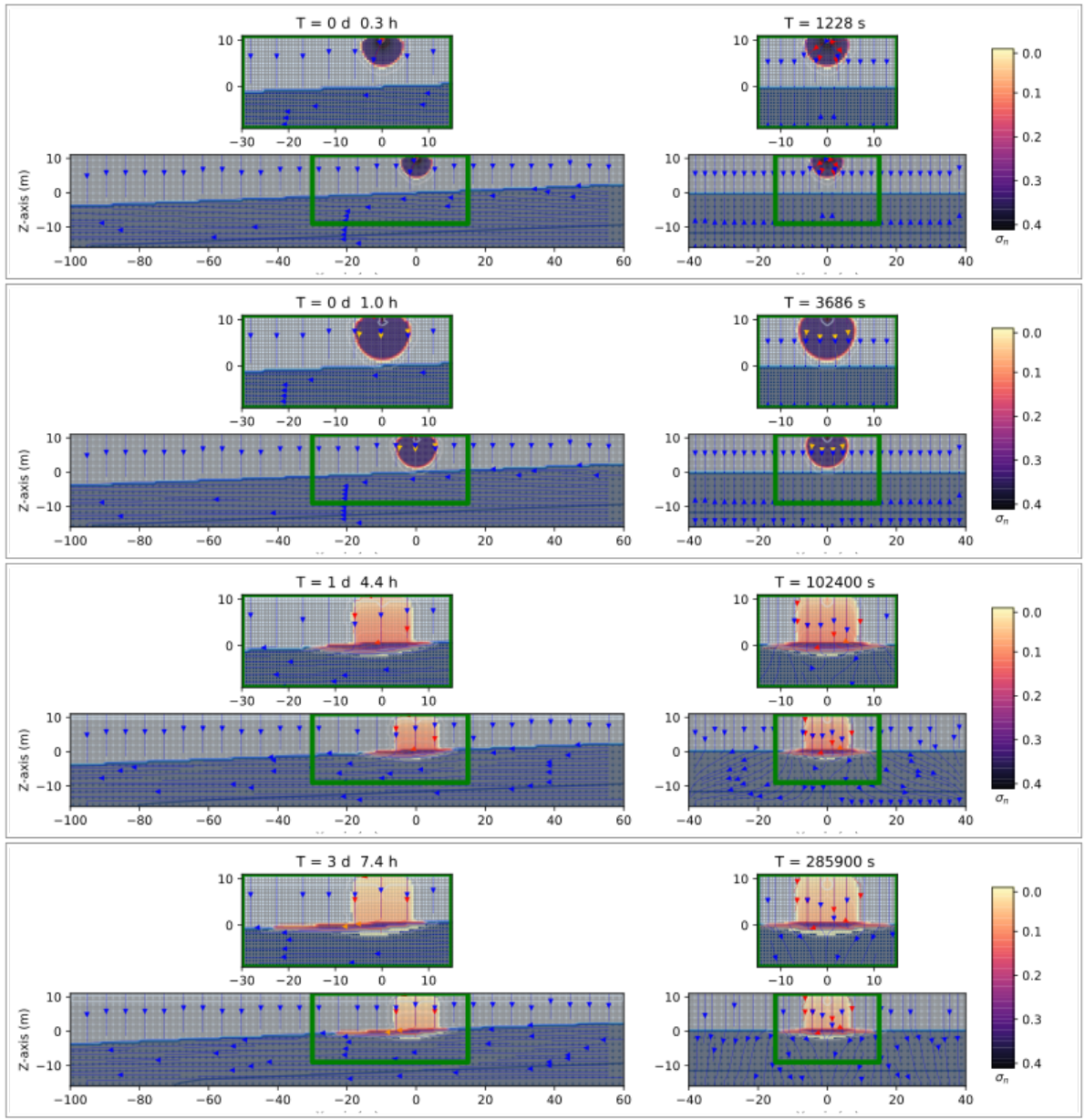

Fig 8. Three-dimensional numerical results on the saturation contours ($\sigma_n = S_n \phi$) of a three-phase immiscible fluid flow (water + gasoline + air) using a spatial grid resolution of 0.50 $m$ and a grid dimension of 160 $m$ × 80 $m$ × 27 $m$, at different times. A hydraulic gradient of 0.04. The spill is released from an oil pipeline at $(x, y, z) = (0,0,10)$ $m$. The unsaturated zone has a $S_w = 0.20$.

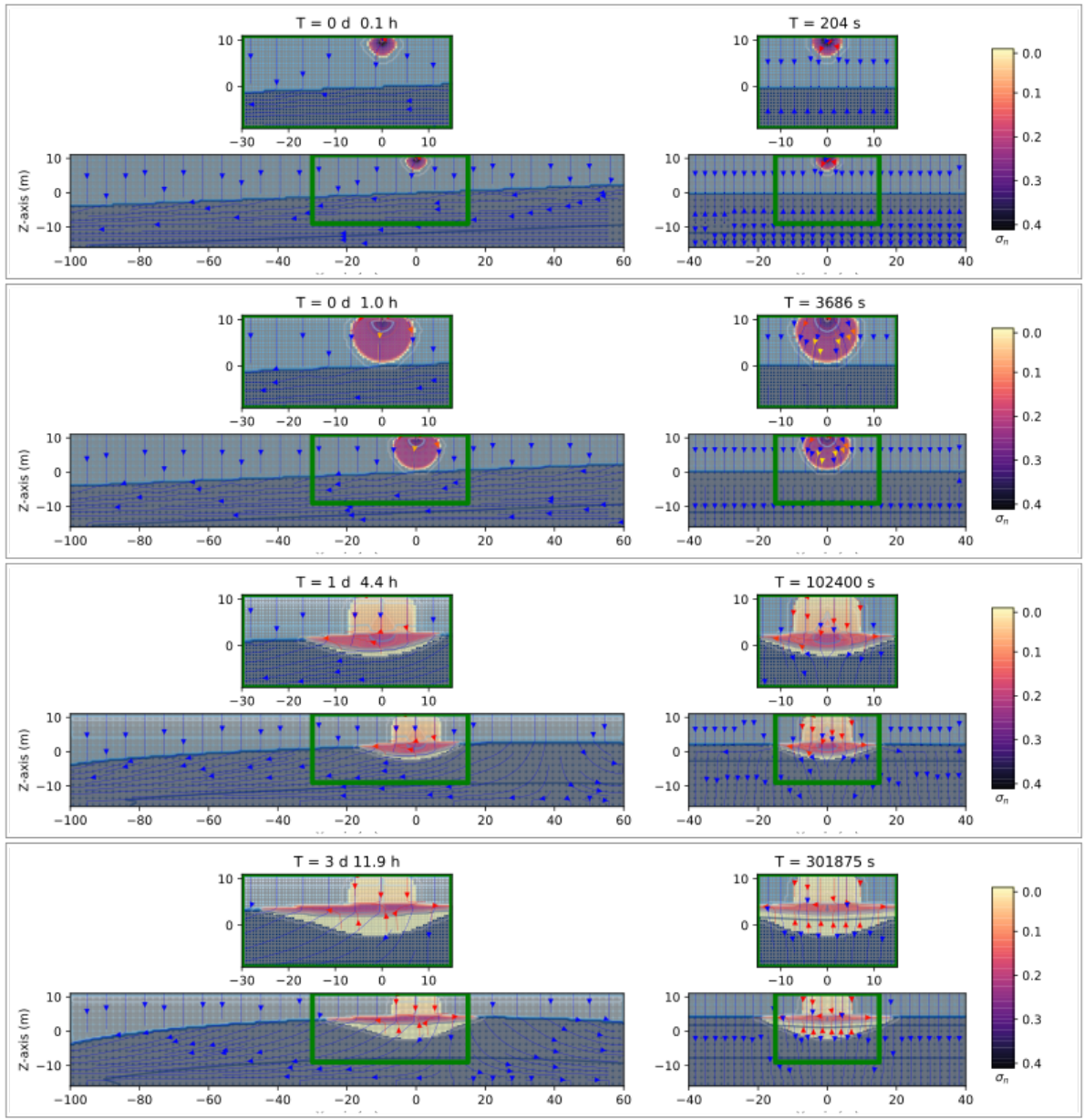

Fig 9. Three-dimensional numerical results on the saturation contours ($\sigma_n = S_n \phi$) of a three-phase immiscible fluid flow (water + gasoline + air) using a spatial grid resolution of 0.50 $m$ and a grid dimension of 160 $m$ × 80 $m$ × 27 $m$, at different times. A hydraulic gradient of 0.04. The spill is released from an oil pipeline at $(x, y, z) = (0,0,10)$ $m$. The unsaturated zone has a $S_w = 0.50$.

## 3.2 Numerical simulations of a diesel oil leak from an oil pipeline

Consider now a leak of diesel oil (denser than gasoline) released into the environment from the oil pipeline situated in the unsaturated zone, just as in the previous cases. The parameters of the numerical simulations are listed in Table 2.

Fig 10 shows similar numerical simulation results as in Fig 2 (dry soil) but using a diesel oil, instead of gasoline of the saturation contours, $(\sigma_n = S_n \phi)$, at different times, for the three-phase immiscible fluid flow (water + diesel oil + air) using a grid resolution of $0.50\ m$ and a grid dimension of $160\ m$ long from $x = [-100, +60]\ m$ (left-hand side), $80\ m$ wide from $y = [-40, +40]\ m$ (right-hand side), and $18\ m$ depth from $z = [+2, -16]m$. The leakage contaminant is initially located at $(x, y, z) = (0,0,1)\ m$ in the oil pipeline (see the first row). After 204 s (the second row), the contaminant arrived at the groundwater table. Once in the groundwater table, it moves to the left-hand side direction due to a hydraulic gradient of 0.04 (the third row). After nine days and 0.8 hours, the diesel oil contaminant reaches position $x = -12.5\ m$ (see left-hand side of the fourth row), in contrast with Fig 2 in which, after nine days and 11.6 hours, the gasoline contaminant reaches position $x = -39.5\ m$.

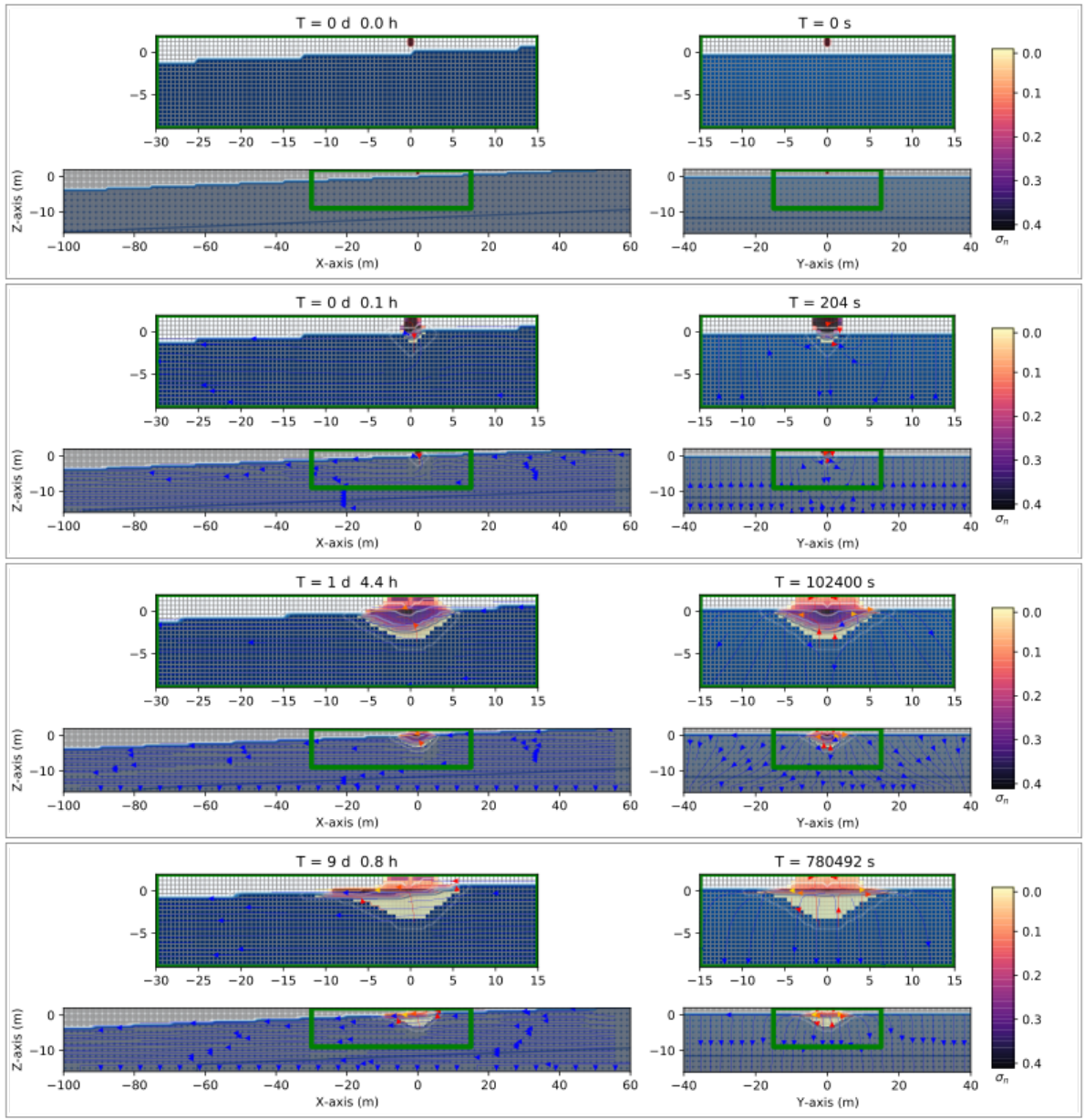

Fig 10. Three-dimensional numerical results on the saturation contours ($\sigma_n = S_n \phi$) of a three-phase immiscible fluid flow (water + diesel oil + air) using a spatial grid resolution of 0.50 $m$ and a grid dimension of 160 $m$ × 80 $m$ × 18 $m$, at different times. A hydraulic gradient of 0.04. The spill is released from an oil pipeline at $(x, y, z) = (0,0,1)\ m$.

Fig 11 shows similar numerical simulation results as in Fig 4 (dry soil) but uses diesel oil instead of gasoline. In Fig 11, the contaminant arrives at the groundwater table long after the gasoline case of Fig 4, say, 3.081 days. In general, it takes longer for diesel oil to arrive at the groundwater table with respect to gasoline and has less mobility to move together with the groundwater flow generated by a hydraulic gradient equal to 0.04.

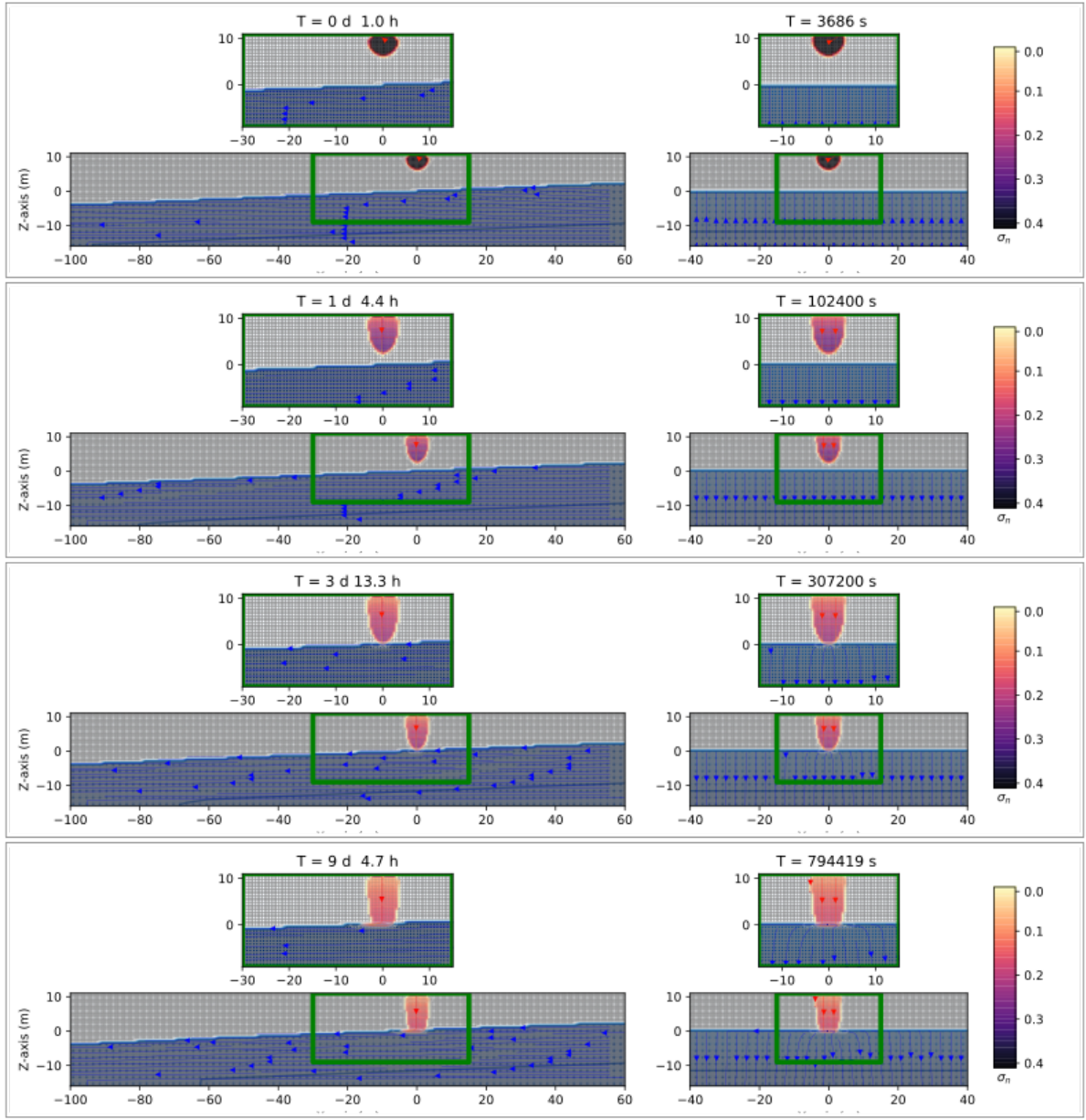

Fig 11. Three-dimensional numerical results on the saturation contours ($\sigma_n = S_n\phi$) of a three-phase immiscible fluid flow (water + diesel oil + air) using a spatial grid resolution of 0.50 $m$ and a grid dimension of 160 $m$ × 80 $m$ × 27 $m$, at different times. A hydraulic gradient of 0.04. The spill is released from an oil pipeline at $(x, y, z) = (0,0,10)$ $m$.

See the Supplementary Material for a similar case with a hydraulic gradient equal to 0.004 instead of 0.04.

### 3.2.1 Numerical simulations of a diesel oil leak from an oil pipeline in a partially unsaturated zone

The effects of the water saturation in the unsaturated zone are also investigated in the case of diesel oil. Consider the two different cases, $S_w = 0.20$ and $S_w = 0.50$, in the unsaturated zone. Fig 12 shows the numerical results on the saturation contours ($\sigma_n = S_n \phi$) of a three-phase immiscible fluid flow (water + diesel oil + air) using a spatial grid resolution of 0.50 $m$ and a grid dimension of 160 $m$ × 80 $m$ × 18 $m$, at different times. A hydraulic gradient of 0.04. The spill is released from an oil pipeline at $(x, y, z) = (0,0,1)\ m$. The unsaturated zone has a $S_w = 0.20$.

After 204 s, the contaminant has already arrived at the groundwater table, as in the case of Fig 10. Notice how the contaminant enters the saturated zone (due to an elevated pressure coming out from the pipeline). Then the contaminant starts to float and moves along with the groundwater flow (see the third row). After five days and 14 hours, the diesel oil reaches position $x = -8.5\ m$ to be compared with the $x = -9.5\ m$ in the case of a diesel oil spill in a dry unsaturated zone (see Fig 10) after five days and 22 hours.

Fig 13 shows the same situation as Fig 12, but with $S_w = 0.50$. As can be seen, the contaminant arrived at the groundwater table at 204 s (the first row), since the effects on the oil pipeline pressure are much stronger than the effects on the saturation of the unsaturated zone. After five days and 14.0 hours (the fourth row) the diesel oil contaminant does not move too much with respect to the previous case with water saturation equals to 0.20 and dry soil. Notice that the level of the groundwater table increases (the fourth row) due to the unsaturated zone being filled with water.

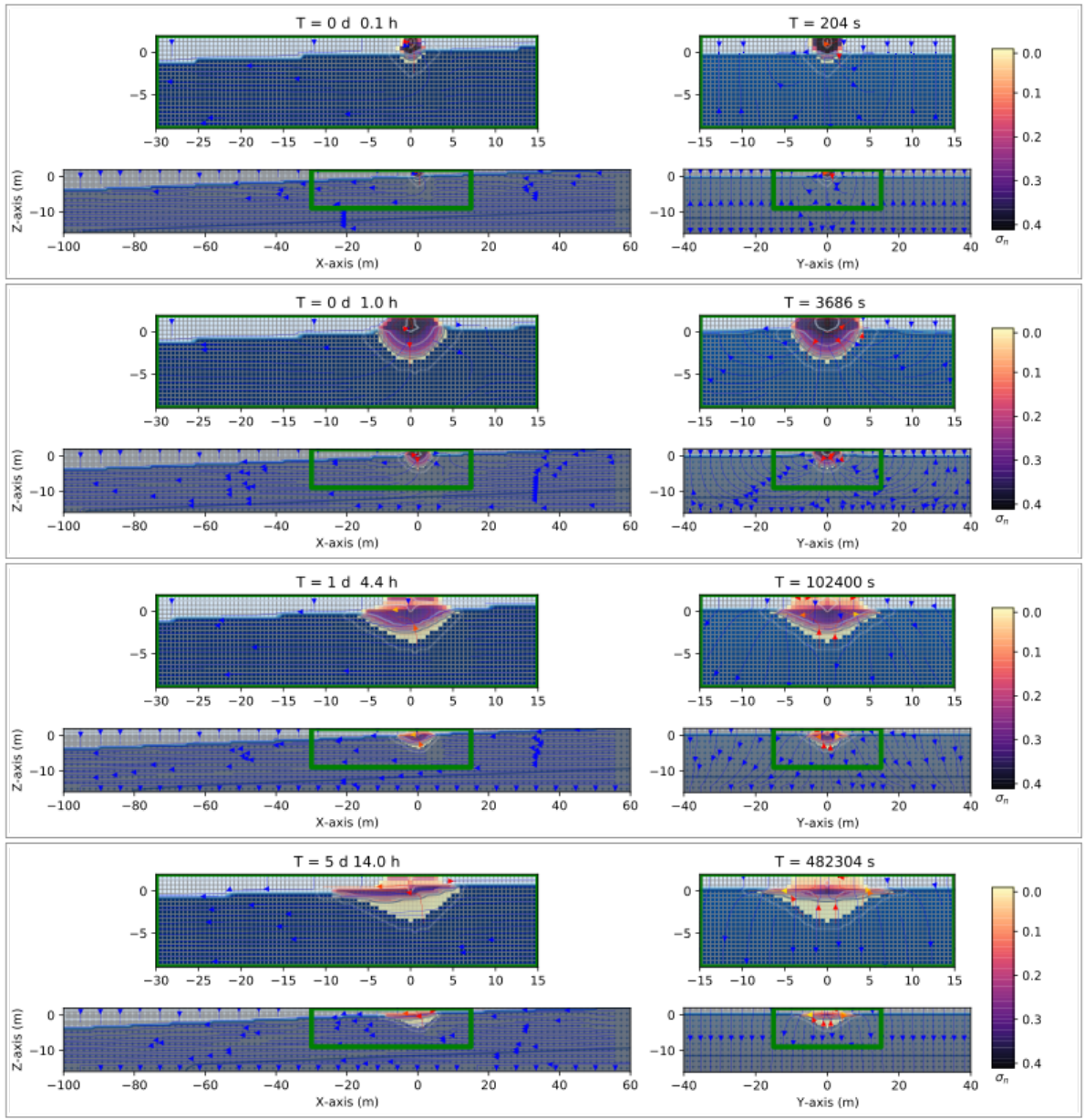

Fig 12. Three-dimensional numerical results on the saturation contours ($\sigma_n = S_n \phi$) of a three-phase immiscible fluid flow (water + diesel oil + air) using a spatial grid resolution of 0.50 $m$ and a grid dimension of 160 $m$ × 80 $m$ × 18 $m$, at different times. A hydraulic gradient of 0.04. The spill is released from an oil pipeline at $(x, y, z) = (0,0,1)$ $m$. The unsaturated zone has a $S_w = 0.20$.

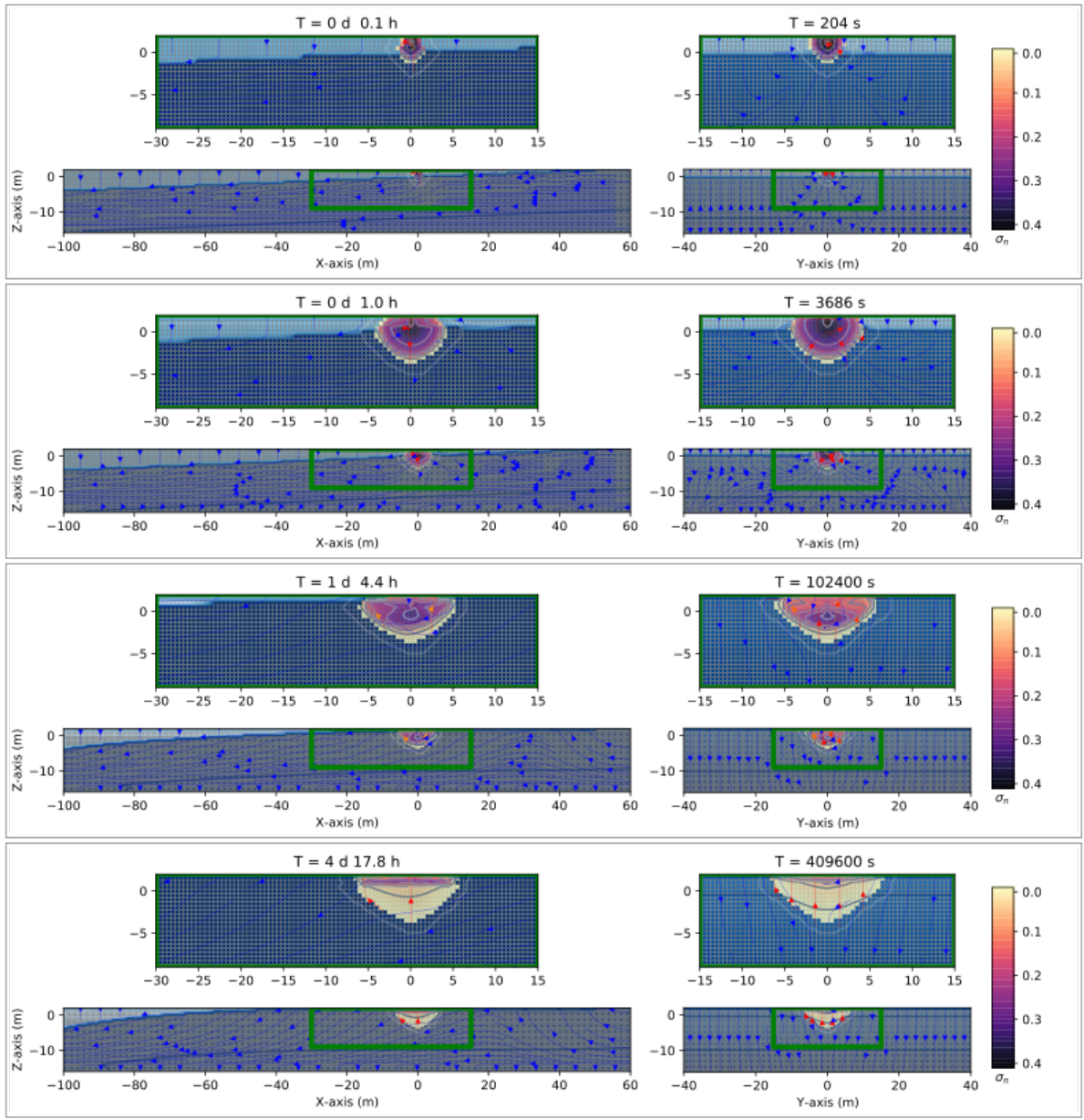

Fig 13. Three-dimensional numerical results on the saturation contours ($\sigma_n = S_n \phi$) of a three-phase immiscible fluid flow (water + diesel oil + air) using a spatial grid resolution of 0.50 $m$ and a grid dimension of 160 $m$ × 80 $m$ × 18 $m$, at different times. A hydraulic gradient of 0.04. The spill is released from an oil pipeline at $(x, y, z) = (0,0,1)\ m$. The unsaturated zone has a $S_w = 0.50$.

Fig 14 and Fig 15 show the same situation as Fig 12 and Fig 13, respectively, except that the diesel oil spill is now released at 10 m instead of one meter from the oil pipeline. The diesel oil contaminant in the case of $S_w = 0.20$, Fig 14, arrives to the groundwater table at 1.754 days, while of the water saturation is $S_w = 0.50$, Fig 15, then the contaminant arrives to the groundwater table at 0.853 days.

Both results to be compared with the similar case but a dry unsaturated zone (Fig 4) in which the diesel oil arrive at the groundwater table at 3.319 days. See the next subsections, for details.

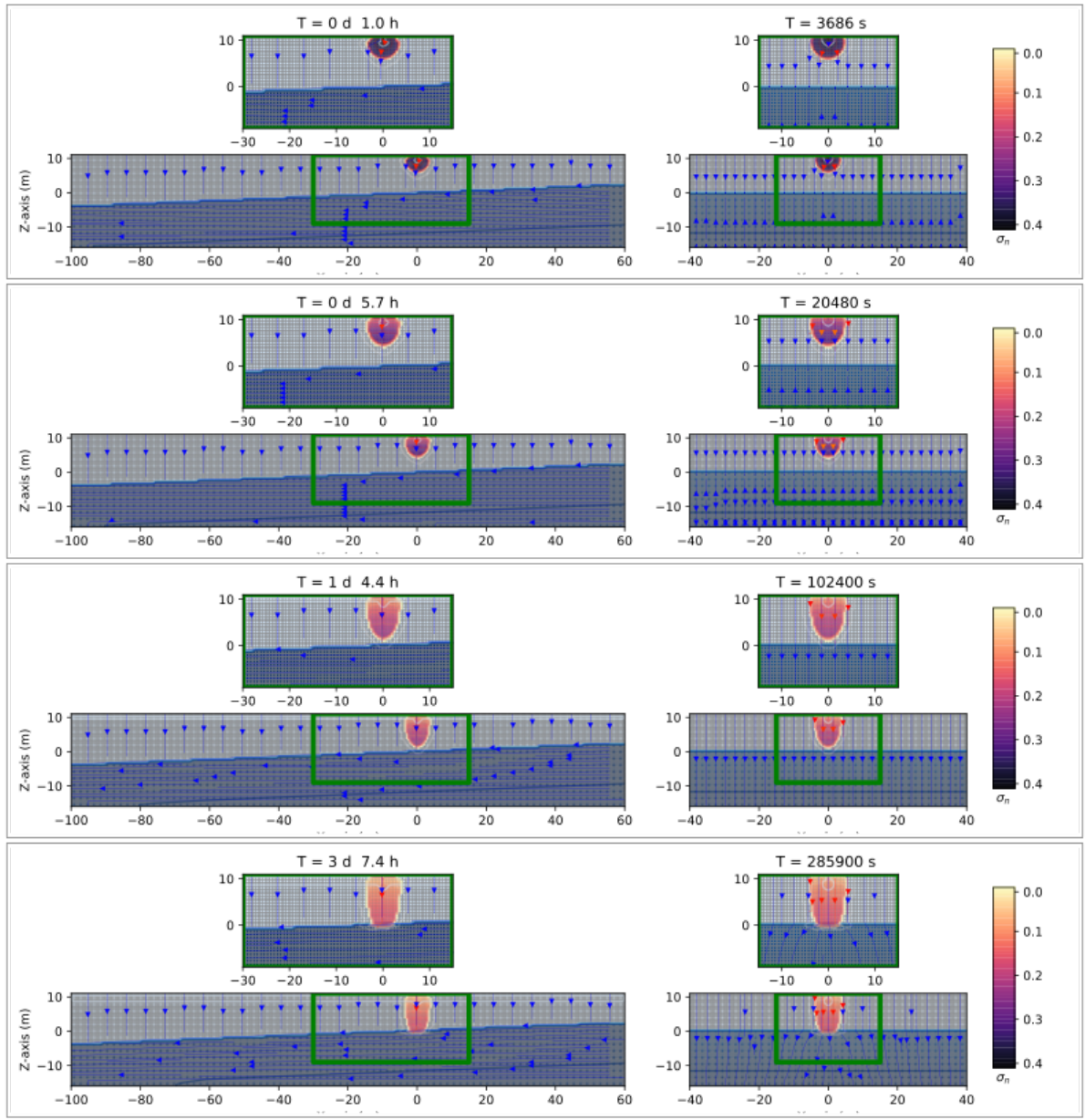

Fig14. Three-dimensional numerical results on the saturation contours ($\sigma_n = S_n \phi$) of a three-phase immiscible fluid flow (water + diesel oil + air) using a spatial grid resolution of 0.50 $m$ and a grid dimension of 160 $m$ × 80 $m$ × 27 $m$, at different times. A hydraulic gradient of 0.04. The spill is released from an oil pipeline at $(x, y, z) = (0,0,10)$ $m$. The unsaturated zone has a $S_w = 0.20$.

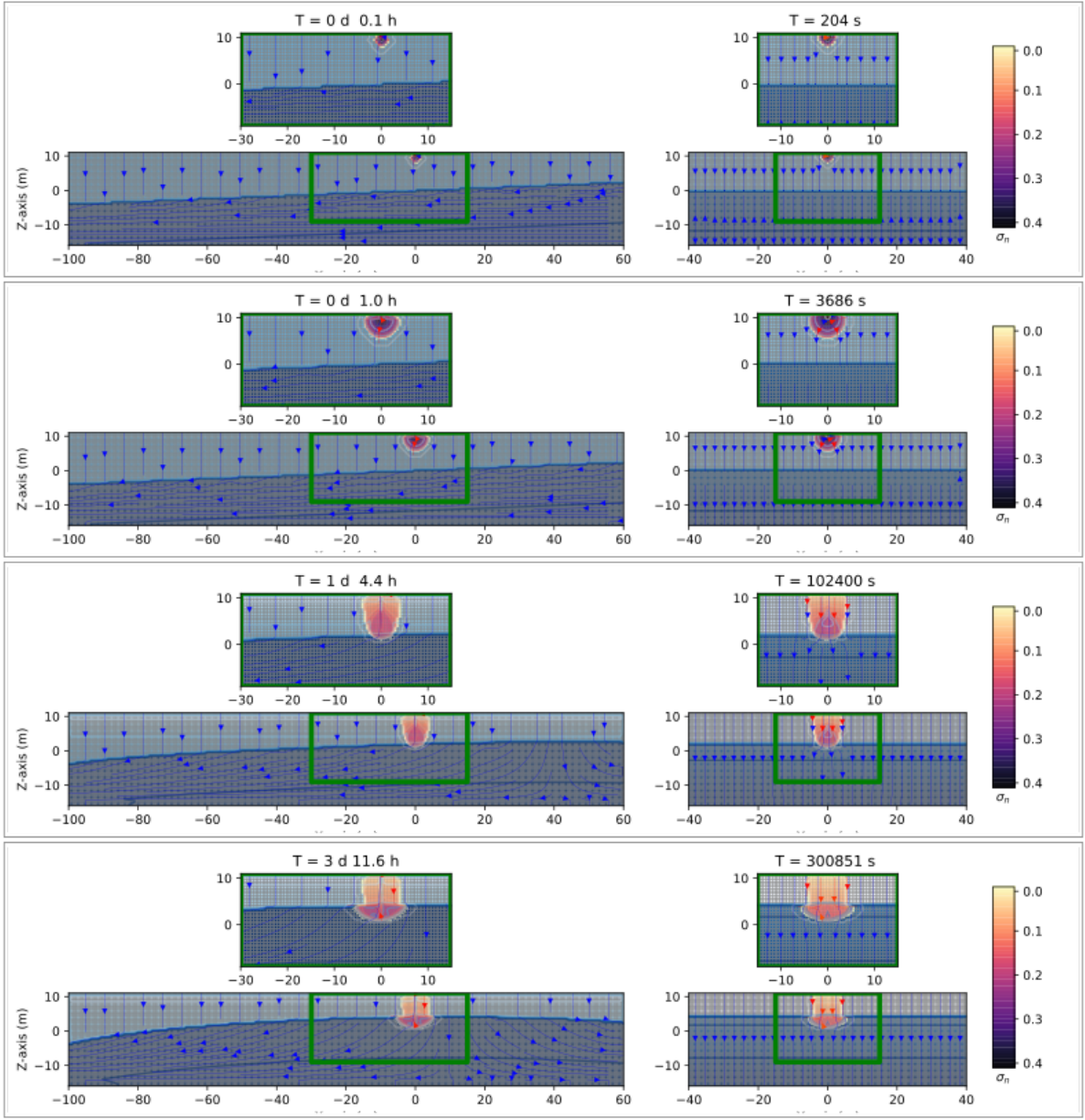

Fig 15. Three-dimensional numerical results on the saturation contours ($\sigma_n = S_n \phi$) of a three-phase immiscible fluid flow (water + diesel oil + air) using a spatial grid resolution of 0.50 $m$ and a grid dimension of 160 $m$ × 80 $m$ × 27 $m$, at different times. A hydraulic gradient of 0.04. The spill is released from an oil pipeline at $(x, y, z) = (0,0,10)$ $m$. The unsaturated zone has a $S_w = 0.50$.

Further numerical simulation results on the saturation contours of a multiphase immiscible fluid flow using either gasoline or diesel oil for 2 m, 5 m, 20 m, and a hydraulic gradient of 0.04 and 0.004 are reported in the Supplementary material.

## 3.3 Effects on the density of the contaminant

Two species of oil (gasoline and diesel oil) with densities of $750 kg/m^3$ and $850 kg/m^3$, respectively, are considered to analyze the influence of oil density on oil spreading in the case of an oil pipeline leaking and a dry soil. Table 3 reported the time employed for the LNAPL contaminant in a numerical simulation to arrive at the groundwater table and the position in the $x$-coordinate at one day and 4.4 hours.

| Type (density) | Thickness of the unsaturated zone (below the pipeline) (m) | Hydraulic gradient | Time to arrival at the groundwater table (s) | Position in -x after 1 day and 4.4 hours (m) |
|---|---|---|---|---|
| Gasoline | 1 | 0.04 | ≤ 204.8 | -16.0 |
|  |  | 0.004 | ≤ 204.8 | -14.0 |
| Diesel oil | 1 | 0.04 | ≤ 204.8 | -6.0 |
|  |  | 0.004 | ≤ 204.8 | -5.5 |
| Gasoline | 2 | 0.04 | ≤ 204.8 | -16.5 |
|  |  | 0.004 | ≤ 204.8 | -15.0 |
| Diesel oil | 2 | 0.04 | 614.4 | -6.0 |
|  |  | 0.004 | 614.4 | -5.0 |
| Gasoline | 5 | 0.04 | 1228.8 | -16.5 |
|  |  | 0.004 | 1228.8 | -15.0 |
| Diesel oil | 5 | 0.04 | 10854.4 | -3.5 |
|  |  | 0.004 | 12288.0 | -3.5 |
| Gasoline | 10 | 0.04 | 6348.8 | -14.0 |
|  |  | 0.004 | 6348.8 | -12.5 |
| Diesel oil | 10 | 0.04 | 286720.0 | x |

|          |    | 0.004 | 276480.0 | x    |
|----------|----|-------|----------|------|
| Gasoline | 20 | 0.04  | 92160.0  | -2.5 |
|          |    | 0.004 | 92160.0  | -2.5 |
| Diesel oil | 20 | 0.04 | x       | x    |
|          |    | 0.004 | x        | x    |

Table 3. Analysis of arrival time at the groundwater table and the position in the -x coordinate under different parameter conditions: two different densities, two different hydraulic gradients, and five different depths from the oil pipeline spill and dry soil. The x sign indicates no available results, i.e., the contaminant is still moving in the unsaturated zone and has not reached the groundwater table.

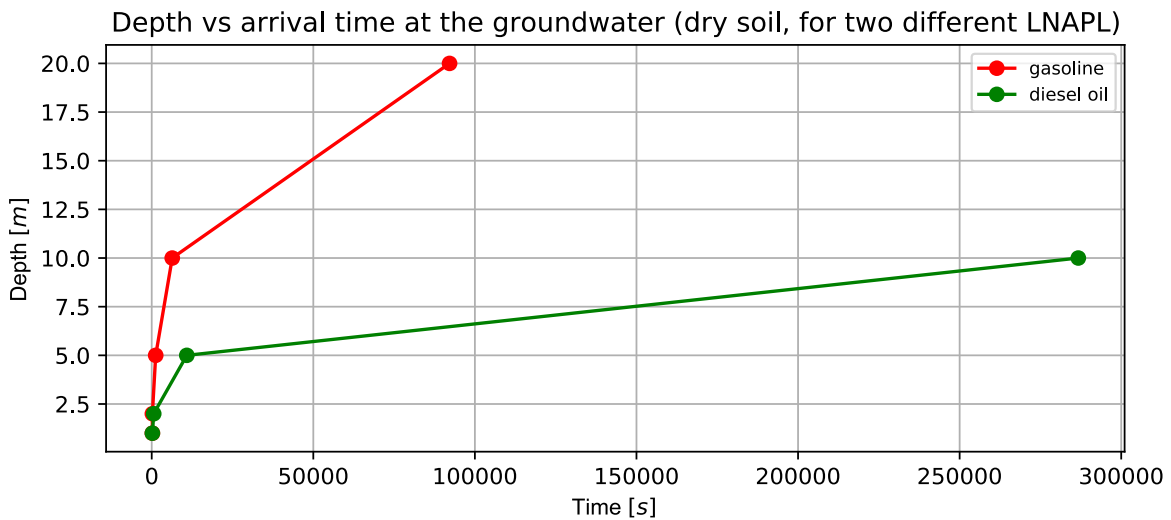

Fig 16. Depth versus arrival time at the groundwater table for dry soil, for two different LNAPLs. The data was taken from the numerical simulations of the saturation contours previously shown on Table 3.

Fig 16 shows the depth of the oil pipeline as a function of the arrival time at the groundwater table in unsaturated dry soil for two different species of oil (gasoline, red line, and diesel oil, green line) and a hydraulic gradient of 0.04. As can be seen from this figure, there are no substantial differences between both species for depths below 2.0 m since the oil spills have a very high pressure with respect to the atmospheric pressure of the unsaturated zone. For this reason, other effects are not noticeable. For depths greater than 5 meters, the time it takes diesel oil to arrive at the groundwater table is much longer than gasoline, although it is much denser. This is because diesel oil's viscosity is greater than gasoline's (see table 1 and table 2). In general, diesel oil spreads less than gasoline once it arrives at the groundwater table. Table 3 also shows that the arrival time to the groundwater table does not depend on the value of the hydraulic gradient in the saturated zone, as it should. But the hydraulic gradient affects the displacement in the x direction. See, for example, the last column of Table 3, where it clearly shows that the displacement is smaller when the hydraulic gradient equals 0.004.

## 3.4 Effects on the water saturation of the unsaturated zone

The effects on the water saturation of the unsaturated zone are reported in Table 4. The time employed for the LNAPL contaminant in a numerical simulation to arrive at the groundwater table and the position in the $x$ coordinate at one day and 4.4 hours are investigated as a function of the depth of the unsaturated zone, and the two different values of the water saturation (hydraulic gradient of 0.04) as were shown in Figs 6-9,12-15.

Fig 17 shows the water saturation of the unsaturated zone as a function of the arrival time at the groundwater table for both gasoline (left-hand side) and diesel oil (right-hand side), reported in Table 4. The data was taken from the numerical simulations. The arrival time at the groundwater table increases as the water saturation goes to zero (dry soil), for both LNAPL densities, although the time for the diesel oil case is much larger than the gasoline case since the viscosity is an order of magnitude higher than the case of the diesel oil (the right-hand side).

Table 3 and Table 4 show the effects on the thickness of the unsaturated zone (the oil pipeline is situated one-meter depth in the unsaturated zone, and the unsaturated zones vary from 1.0 up to 20.0 m. The more profound the unsaturated zone is, the more time the contaminant takes to arrive at the groundwater table.

| Depth of the unsaturated zone | Type of contaminant | Water saturation in the | | Position in -x after 1 day |
|---|---|---|---|---|

| (below the pipeline) (m) | | unsaturated zone | Time of arrival to the groundwater table (s) | and 4.4 hours (m) |
|---|---|---|---|---|
| 1.0 | gasoline | 0.0 | ≤ 204.8 | -16.0 |
| | | 0.2 | ≤ 204.8 | -16.0 |
| | | 0.5 | ≤ 204.8 | -16.0 |
| 1.0 | diesel oil | 0.0 | ≤ 204.8 | -6.0 |
| | | 0.2 | ≤204.8 | -6.0 |
| | | 0.5 | ≤204.8 | -6.0 |
| 10.0 | gasoline | 0.0 | 6348.8 | -14.0 |
| | | 0.2 | 4915.0 | -16.0 |
| | | 0.5 | 4710.4 | -16.0 |
| 10.0 | diesel oil | 0.0 | 286720.0 | x |
| | | 0.2 | 151552.0 | x |
| | | 0.5 | 73720.0 | -1.5 |

Table 4. Analysis of arrival time at the groundwater table and the position in the -x coordinate under different parameter conditions: two different types of densities, two different depths from the oil pipeline spill and, water saturation of the unsaturated zone equal to 0.20 and 0.50. The hydraulic gradient is 0.04.

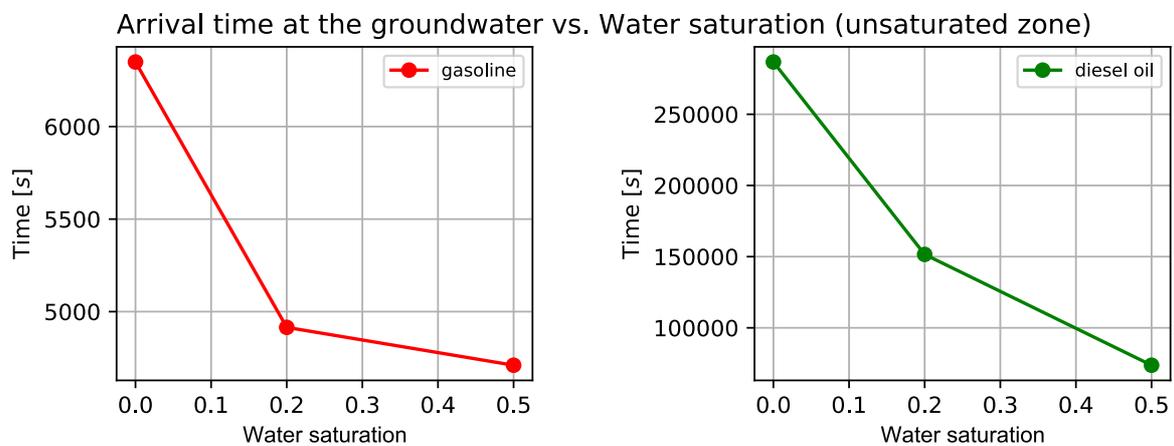

Arrival time at the groundwater vs. Water saturation (unsaturated zone)

Fig 17. Water saturation in the unsaturated zone versus arrival time at the groundwater table, for two different species of LNAPL. Red line corresponds to gasoline and red line corresponds to diesel oil.

## 3.5 Effects on the pressure in the oil pipeline

All the simulations previously shown consider an oil pipeline pressure of 2.0265 MPascal during the time of the leakage, which was fixed to be one hour, and an atmospheric pressure value in the unsaturated zone as the initial condition. The high pressure inside the oil pipeline has the most critical effect on the arrival time of the contaminant at the groundwater table when the distance between the leakage and the groundwater table is smaller than two meters, as was previously shown. Consider now the case in which the pressure inside the oil pipeline is also atmospheric. Table 5 shows the numerical simulation data results for the arrival time for the gasoline spill with an atmospheric pressure, different values for the water saturation in the unsaturated zone, and a hydraulic gradient of 0.04.

| Depth of the unsaturated zone (below the pipeline) (m) | Type of contaminant | Water saturation in the unsaturated zone | Arrival time at the groundwater table (s) |
|---|---|---|---|
| 1.0 | gasoline | 0.0 | 1024.0 |
| 1.0 | gasoline | 0.20 | 819.2 |
| 1.0 | gasoline | 0.50 | 614.4 |

Table 5. Water saturation of the unsaturated zone vs. arrival time at the groundwater table for a gasoline spill at the atmospheric pressure.

Fig 18 shows the arrival time of the gasoline spill at the groundwater table as a function of the water saturation in the unsaturated zone at one-meter depth (see Table 5) and the case in which the contaminant spill has an atmospheric pressure (green line) and another case in which the spill pressure is $2.0 \times 10^6 Pa$ (red line, previous cases). As can be seen, in the case of high pressure, the water saturation of the unsaturated zone does not affect the arrival time very much since the depth is not much longer (red line), and the effect of the high pressure is stronger. In the case of a gasoline spill with atmospheric pressure, the arrival time to the groundwater table decreases when the water saturation of the unsaturated zone increases.

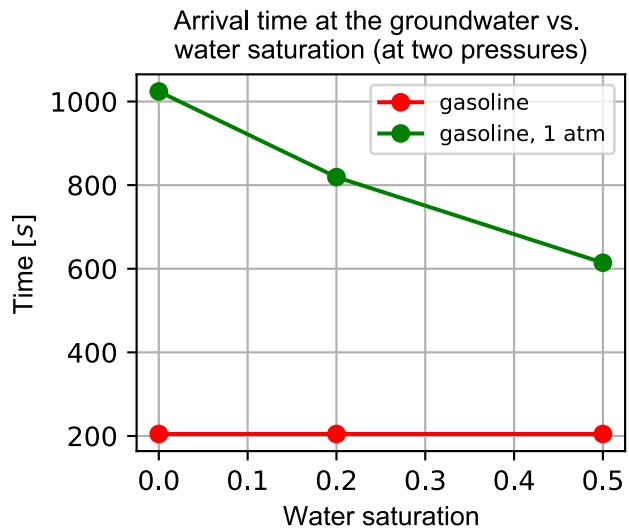

Fig 18. Arrival time at the groundwater table vs. water saturation of the unsaturated zone for two different pressure spills of the gasoline.

## 3.6 Model Validation (refinement)

In a previous publication (Feo and Celico, 2021,2022), the numerical method (HRSC) and the code was validated using several toy models but also a real experiment such as a sandbox. Since here, it was not possible to validate the numerical results to experimental ones, we had to rely on classical convergence test running the same code at different resolutions. Fig 19 shows comparison results on the saturation contours for a gasoline contaminant spill from an oil pipeline at two different times using a grid resolution of 0.25 m and a time step size of 0.00625 s (first and third row) and a grid resolution 0.50 m and a time step size of 0.0025 s (second and fourth row, see Fig 2). Both results are similar and show that using two different resolutions the results are similar within a 5% error.

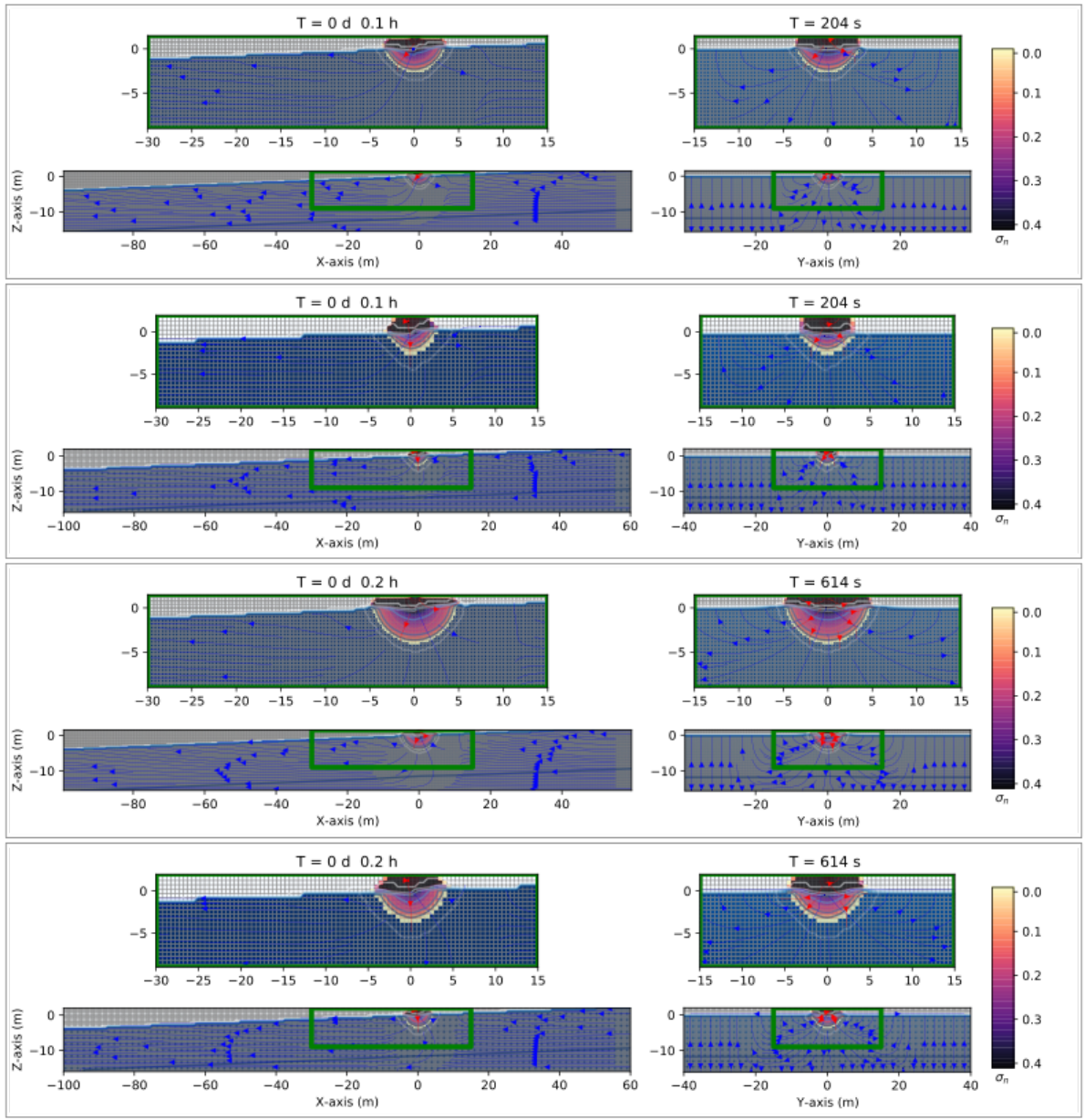

Fig 19. Saturation contours ($\sigma_n = S_n \phi$) of a three-phase immiscible fluid flow (water + gasoline + air) using a grid resolution of 0.25 m (first and third row) and a grid resolution 0.50 m (second and fourth row, see Fig 2) at two different times in the case of a gasoline spill from the oil pipeline.

## 4. Conclusions

In this work, we presented a high-resolution three-dimensional numerical modeling investigation on the effects of an oil pipeline failure in a range of hydrogeological conditions, using high-resolution shock-capturing (HRSC) flux conservative method, and the CactusHydro code, recently introduced in (Feo and Celico, 2021, 2022). We investigated two different density types (gasoline and diesel oil) in a variably saturated zone and the effects on the variation of the depth of the unsaturated zone, three different saturations of the unsaturated zone that include the dry soil case, the hydraulic gradient and the pressure of the oil pipeline leakage. We investigate the contours saturation of the contaminant and its migration from the oil pipeline leakage with great accuracy. The results indicate that the leaking oil's pressure is the parameter that most significantly affects the contaminant's arrival at the groundwater table. Also, the water saturation of the unsaturated zone influences the arrival time as the water saturation increases, the time of arrival decreases (for a fixed depth). This is due to the fact that the contaminant moves more easiest downward when the saturation increases (Acher et al., 1989). The unsaturated depth zone significantly influences the contaminant migration in the saturated zone/capillary fringe, although it results no important in the vertical direction. The unsaturated zone depth significantly influences the contaminant migration unsaturated zone, while the oil density and the hydraulic gradient have limited effects on the contaminant migration in the variably saturated zone.

In the application field, CactusHydro code represents an advanced tool to be applied in the environmental risk assessment caused by hydrocarbon releases by onshore pipelines, in accordance with the provisions of the Legislative Decree 26 June 2015, n. 105 "Implementation of Directive 2012/18/EU relating to the control of the danger of major accidents connected with dangerous substances" (the so-called Seveso III Directive). The assessment of the aquifer vulnerability and the analysis of the environmental consequences of a major accident are in fact specific components of the Safety Reports that must be prepared by the Plant Managers as required by the aforementioned legislation. With this in mind, the results of the application of CactusHydro code can be used as a support of utmost importance for identifying the most effective actions aimed at preventing and/or reducing the probability and extent of "pollution" and damage to environmental receptors in the event of major accident. In conclusion, it should be emphasized that the CactusHydro code can also be used at 360 degrees to evaluate the effects and environmental impacts deriving from leaks of hydrocarbons and toxic substances, i.e., dense nonaqueous phase liquid from underground or above-ground tanks.

# Appendix A. Supplementary material

Supplementary material related to this article can be found, in the online version, at doi xx:

# Acknowledgements

This work used resources at the University of Parma at (https://www.hpc.unipr.it). This work has benefit from the equipment and framework of the COMP-HUB Initiative, funded by the "Departments of Excellence" program of the Italian Ministry of Education, University and Research (MIUR, 2018-2022).